\documentstyle[aps,eqsecnum,preprint,epsf
]{revtex}
\begin{document}

\tighten
\preprint{ETH-TH/98-19}
\draft
\title{Amplification of Cosmological Inhomogeneities by the QCD Transition}
\author{Christoph Schmid$^1$\thanks{e-mail: chschmid@itp.phys.ethz.ch}, 
    Dominik J. Schwarz$^2$\thanks{e-mail: dschwarz@th.physik.uni-frankfurt.de},
    and Peter Widerin$^1$\thanks{e-mail: widerin@itp.phys.ethz.ch}}
\address{$^1$Institut f\"ur Theoretische Physik, 
         ETH-H\"onggerberg, 8093 Z\"urich, Switzerland\\
         $^2$Institut f\"ur Theoretische Physik, Universit\"at Frankfurt,\\
         Postfach 11 19 32, 60054 Frankfurt am Main, Germany}
\maketitle

\begin{abstract}
The cosmological QCD transition affects primordial density perturbations. 
If the QCD transition is first order,
the sound speed vanishes during the transition and density perturbations 
fall freely. For scales below the Hubble radius at the transition
the primordial Harrison-Zel'dovich spectrum of density fluctuations  
develops large peaks and dips. These peaks grow with wave number for both 
the hadron-photon-lepton fluid and for cold dark matter. At the horizon scale
the enhancement is small. This by itself does not lead to the formation
of black holes at the QCD transition. The peaks in the hadron-photon-lepton
fluid are wiped out during neutrino decoupling. For cold dark matter
that is kinetically decoupled at the QCD transition (e.g., axions or
primordial black holes) these peaks lead to the formation of CDM clumps of
masses $10^{-20} M_\odot< M_{\rm clump} < 10^{-10} M_\odot$.
\end{abstract}

\section{Introduction and Results}

In the early universe a transition from a quark-gluon plasma (QGP) to a
hadron gas (HG) took place at a certain transition temperature $T_\star$.
At high temperatures asymptotic freedom of QCD predicts the existence of a
deconfined phase, the quark-gluon plasma. At low temperatures quarks and
gluons are confined in hadrons. The deconfined phase at high temperatures
was found by lattice QCD simulations \cite{lattice}. The order of the QCD
transition is still under debate, but recent lattice QCD results for
three quark flavors with massive strange quarks indicate that the QCD
transition is first order \cite{Iwasaki}. The value of the latent heat is
available only for quenched lattice QCD (gluons only, no quarks)
\cite{Iwasaki2}. These data give a small latent heat $l$; $(l/T_\star)$ is
about one fifth of the difference in entropies between an ideal massless
quark-gluon plasma resp. hadron gas. Two-flavor calculations give
$T_\star\approx140$ MeV \cite{MILC96}. Ongoing and future heavy ion
experiments at the world's largest colliders (SPS, RHIC, LHC) are searching
for the QGP \cite{search}.

At the QCD transition the Hubble radius $R_{\rm H} \sim m_{\rm Pl}/T^2_\star$ 
is about $10$ km. The mass inside a Hubble volume is $\sim 1 M_\odot$.
The transition lasts about a tenth of a Hubble time, $t_{\rm H}$, with the
latent heat from quenched lattice QCD. The Hubble
time at the QCD transition is $t_{\rm H} = R_{\rm H}/c \sim 10^{-5}$ s. 
This is extremely long in comparison with the relaxation time scale of the
strong interactions, which is about $10^{-23}$ s. Thus, the 
transition is very close to an equilibrium process. 

For a first-order QCD transition hadronic bubbles nucleate during a short 
period of supercooling. The crucial parameters for supercooling are the
surface tension $\sigma$ and the latent heat $l$. Data for $\sigma$ and $l$
are available only for quenched QCD \cite{Iwasaki2} and give a very small
surface tension. This gives a very short period of supercooling, $\Delta
t_{\rm sc} \sim 10^{-3} t_{\rm H}$, and a typical bubble nucleation distance 
$d_{\rm nucl}$ $\approx 1$ cm $\approx 10^{-6} R_{\rm H}$ for homogeneous 
nucleation \cite{Ignatius,Christiansen}. The hadronic bubbles grow very
fast, within $10^{-6} t_{\rm H}$, until the released latent heat has reheated
the universe to $T_\star$. For the remaining $99\%$ of the transition the HG
and the QGP coexist at the pressure $p_{\rm HG}(T_\star) = p_{\rm
QGP}(T_\star)$. During this time the hadronic bubbles grow slowly and the
released latent heat keeps the temperature constant until the transition is
completed. The energy density decreases continuously from $\rho_{\rm
QGP}(T_\star)$ at the beginning of the transition to $\rho_{\rm HG}(T_\star)$ 
when the transition is completed.

In the mid 80s interest in the cosmological QCD transition arose because
it was realized that a strongly first-order QCD phase transition with
large surface tension  could lead to observable signatures today. Witten
\cite{Witten} pointed out that the separation of phases during 
the coexistence of the hadronic and the quark phase could gather most 
baryons in (strange) quark nuggets. 
However, the quark nuggets, while cooling, lose baryons, and evaporize, 
unless they contain much more than $10^{44}$ baryons initially 
\cite{evap}. This number should be compared with the number of
baryons inside a Hubble volume at the QCD transition, which is
$10^{50}$. Thus, the mean bubble nucleation distance should be $> 10^{-2}
R_{\rm H} \sim 100$ m, which is about a factor $10^4$ too big compared to
the nucleation distance suggested using recent lattice results
\cite{Iwasaki2}.   
 
Applegate and Hogan found that a strongly first-order QCD phase
transition with large surface tension could induce inhomogeneous
nucleosynthesis \cite{BBN}, assuming a nucleation distance which is ruled out
by recent lattice results \cite{Iwasaki2}. The large isothermal baryon
fluctuations, induced during the separation of phases,  could lead to
inhomogeneous initial conditions for nucleosynthesis. As a minimal
requirement for inhomogeneous nucleosynthesis the mean bubble nucleation
distance should be larger than the proton diffusion length, which
corresponds to $\sim 3$ m \cite{Mathews} at the QCD transition. This is more
than $10^2$ times larger than the nucleation distance based on lattice
results  \cite{Iwasaki2}. In addition, the observed cosmic abundances of light
elements do not allow inhomogeneous nucleosynthesis, except in a small region
in parameter space corresponding to an inhomogeneity scale of $\sim 40$ m
\cite{Mathews}.     

In a recent letter \cite{SSW} and in the present work we look at matter
averaged over scales $\lambda$ much larger than the bubble separation,
e.g. $\lambda$ of order of the Hubble radius $R_{\rm H}$. Therefore we deal
with the bulk properties of the fluid in contrast to previous investigations
which dealt with physics at the bubble separation scale. We found that the
evolution of cosmological density perturbations is strongly affected by a
first-order QCD transition for subhorizon scales, $\lambda < R_{\rm
H}$. Preexisting cosmological perturbations with an approximate
Harrison-Zel'dovich spectrum \cite{HZ} are predicted by inflation
\cite{perturbations,BST,Bardeen,Mukhanov} and have been observed in the
temperature fluctuations of the cosmic microwave background by the COBE
satellite \cite{Smoot}.  

We showed in \cite{SSW} that pressure gradients and the isentropic sound
speed (for wavelengths $\lambda$ much larger than the bubble separation),
$c_s=\left(\partial p/\partial \rho \right)_S^{1/2}$, must be zero during a
first-order phase transition of a fluid with negligible chemical potential
(i.e. no relevant conserved quantum number). The sound speed must be zero,
because for such a fluid the pressure can only depend on the temperature,
$p(T)$, and because the transition temperature $T_{\star}$ has a given value,
it cannot depend on any parameter, hence $p(T_{\star}) = p_{\star}$ is a given
constant, and  $c_s=0$. This is in contrast e.g. to the water-vapor system,
where the number of molecules is conserved, the pressure also depends on the
particle density in a given phase, $p(T,n)$, and where a two-phase system has
an equilibrium pressure (the vapor pressure) which depends on the temperature,
$p_{\star}(T)$, hence $c_s \ne 0$. During the entire QCD transition the sound
speed stays zero and suddenly rises back to the radiation value
$c_s=1/\sqrt{3}$ after the transition is completed. Pressure varies
continuously and goes below the ideal radiation fluid value $p=\rho/3$, but
stays positive. Jedamzik \cite{Jedamzik} independently pointed out that a
significant decrease in the effective sound speed $c_s$ during the
cosmological QCD transition is expected. 

Since the sound speed is zero during a first-order QCD transition, there
are no pressure gradients and no restoring forces. Preexisting cosmological
perturbations go into free fall. The superhorizon modes (at the time of the
transition), $\lambda>R_{\rm H}$, remain unaffected. The spectrum of
subhorizon perturbations develops peaks in $\delta \rho/\rho$ which grow with
the wave number. The details of the enhancement depend on the QCD equation of
state near $T_{\star}$. The peaks grow at most linearly in wave number. The
subhorizon peaks arise because preexisting acoustic oscillations at shorter
wavelengths have a higher velocity (for a given $\delta \rho/\rho$) at the
beginning of the transition and therefore get a larger amplification factor
for $\delta \rho/\rho$ during the transition.
 
Photons and relativistic leptons, i.e. the three neutrinos, $e^\pm$ and
$\mu^\pm$, also contribute to the energy density at the temperatures of
interest. Weak and electro-magnetic interactions are fast enough to keep the
photons and leptons in thermal equilibrium with the
QGP/HG. Therefore the collective behavior of all these coupled particles can
be described by one perfect (i.e. dissipationless) fluid, which we shall
denote as radiation fluid (RAD).   

As a second fluid we include cold dark matter (CDM) which dominates the energy
density of the universe today. At the time of the QCD  
transition $\rho^{\rm CDM}(T_\star) \sim 10^{-8} \rho^{\rm RAD}(T_\star)$,
therefore the gravity generated by CDM can be neglected. We must distinguish
two types of CDM with respect to the issue of kinetic decoupling at the time
of the QCD transition. The lightest supersymmetric particle, which is probably
the neutralino \cite{Griest}, is tightly coupled to the radiation fluid (and
is treated as part of the RAD fluid) at $T \sim T_\star$. The scattering rate
of the neutralino is of the order of the weak interaction scattering rate, the
neutralino decouples kinetically at a much lower temperature of a few
MeV. Kinetic decoupling must be distinguished from freeze out (when the pair
production and annihilation rates fall below the Hubble rate), which happens
way before the QCD transition. In contrast the second type of CDM is
kinetically decoupled at the QCD transition, e.g. axions or preexisting
primordial black holes. Note that in our figures and equations kinetically
decoupled CDM will be labeled by CDM for brevity.

Kinetically decoupled CDM falls into the potential wells provided by the
dominant radiation fluid. Thus, the CDM spectrum is amplified on subhorizon
scales. The peaks in the CDM spectrum go nonlinear shortly after
equality. This leads to the formation of CDM clumps with mass $< 10^{-10}
M_\odot$. Especially the clumping of axions has important implications for the
axion searches using the magnetic field induced axion decay \cite{Sikivie}.  

The formation of primordial black holes (PBHs) should be particularly
efficient during the QCD epoch due to a substantial reduction of pressure
forces during adiabatic collapse as pointed out by Jedamzik \cite{Jedamzik}.
The PBH mass function is expected to exhibit a pronounced peak on the
QCD-horizon mass scale $\sim 1 M_{\odot}$. He proposes that these black holes
could account for massive compact halo objects (MACHOs) observed by
microlensing \cite{microlensing}. However, we found by our linear analysis
\cite{SSW} that the amplification of fluctuations at the QCD horizon crossing
scale is only a factor $1.5$. For standard models of structure formation
without tilt, the amplitudes are not big enough to produce a cosmologically
relevant amount of black holes \cite{COSMO}. A tilted spectrum could be
fine tuned to produce black holes at the QCD scale, but the spectrum would
need a break just below the QCD scale in order not to overproduce smaller
black holes. With the need of such a doubly fine-tuned preexisting spectrum
the main effect would not be due to the QCD transition. 
 
The influence of the QCD transition on primordial gravitational waves
has been investigated by one of the present authors in Ref.~\cite{Schwarz}.
It was found that the dramatic drop in relativistic degrees of freedom
during the QCD transition (from $51.25$ in the QGP to $17.25$ in the HG)
induces a step by 30\% in the primordial spectrum of gravitational
waves. Today this step is located at $\sim 10^{-8}$ Hz. Gravitational 
waves with that frequency could be detected by pulsar timing, 
however for gravitational waves from inflation the amplitudes are far too 
small to show up in the pulsar timing residuals. 

This paper is organized as follows: A discussion of hot QCD, bubble
nucleation, and the vanishing of the sound speed for wavelengths $\lambda$ much
larger than the bubble separation can be found in Sec.~II. In this section
we present a fit to recent lattice QCD results. We give the thermodynamical
curves for $w=p/\rho$ and the sound speed $c_s^2$ as a function of the scale
factor at the QCD transition for three cases: The bag model, \cite{bag}, which
gives a simple parameterization, our fit to lattice QCD and a smooth
crossover.

In Sec.~III we derive the amplification of density perturbations at the QCD
transition both numerically and analytically. We present the evolution
equations for cosmological perturbations in a first-order form in uniform
expansion gauge, which gives the same simple structure for the general
relativistic dynamics as for the Jeans equations. The details of the
amplification depend on the equation of state at the phase transition. We
present three cases: First, we use the bag model \cite{bag}, which allows a
simple discussion of the effects. In the bag model, the amplification of
subhorizon perturbations grows linearly in wave number $k$. Next, we use our
fit to recent lattice QCD data \cite{Boyd,MILC97}, which indicate a smaller
latent heat \cite{Iwasaki2}. The amplifications for the lattice fit grow
proportional to $k^{3/4}$. As a third example we study the possibility of a
smooth crossover \cite{Blaizot}. In this case, the sound speed stays finite
and subhorizon scales are not amplified. The spectrum for a crossover has
peaks only around the horizon scale. We prove that the spectra of primordial perturbations are not affected for scales
greater than the Hubble horizon $R_{\rm H}$ at the transition.  

In the last section (Sec.~IV) we discuss observable consequences of the large
peaks and dips in the spectrum of density fluctuations. Collisional damping
around the time of neutrino decoupling wipes out all subhorizon
inhomogeneities in the radiation fluid before big bang nucleosynthesis (BBN),
and homogeneous BBN cannot be affected. We discuss black hole formation during
the QCD transition and conclude that it is highly unlikely unless the
primordial spectrum is fine tuned at the QCD scale. 

The most interesting prediction is the clumping of those types of CDM which
are kinetically decoupled at the QCD transition and do not suffer from
collisional damping by neutrinos. E.g. axions, if they are the CDM, are not
distributed smoothly  within the halo of our galaxy, they come in clumps. 

Throughout the paper we assume that our universe is spatially flat 
($\Omega = 1$) and most of its matter is CDM today. Whenever we give 
numbers we use a Hubble parameter of $H_0 = 50\; {\rm km}\;{\rm  s}^{-1}{\rm Mpc}^{-1}$.

\section{The cosmological QCD transition}

\subsection{Hot QCD}

The baryon number density in the universe is extremely small compared 
to the entropy density, i.e.~$n_B/s ={\cal O}(10^{-10})$. At the time of 
the QCD transition the chemical potential for quarks resp. baryons is 
negligible as far as the pressure $p$ and energy density $\rho$ are 
concerned, $\rho$ and $p$ depend on the temperature only. 

\begin{figure}
\begin{center}
\epsfig{figure=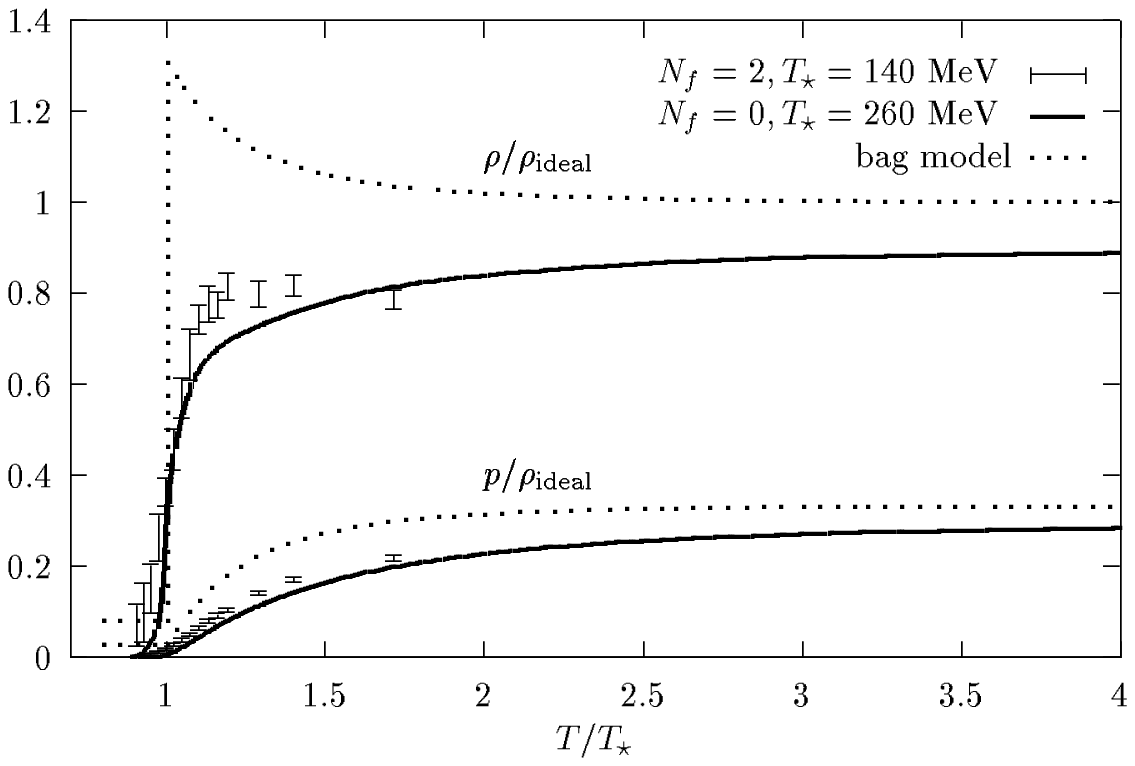,width=0.67\linewidth}
\end{center}
\vspace{-7pt}
\caption{\label{fig1} 
The energy density and the pressure of hot QCD relative to the energy density
of an ideal quark-gluon plasma are plotted as a function of $T/T_\star$.
The lattice QCD data for $N_f = 0$ \protect\cite{Boyd} and 
$N_f = 2$ QCD \protect\cite{MILC96} 
are compared with the predictions of the bag model.}
\end{figure}

The behavior of $\rho(T)$ and $p(T)$ near the QCD transition must be given by
nonperturbative methods, lattice QCD. In Fig.~\ref{fig1} we plot lattice QCD
data for $\rho(T)$ and $p(T)$ divided by $\rho$ of the corresponding ideal
gas. We show the lattice results for two systems, quenched QCD (no quarks)
\cite{Boyd}, and two-flavor QCD \cite{MILC97}. For quenched QCD the lattice
continuum limit is shown. For two flavor QCD the data with six time steps
($N_t = 6$, $a\approx 0.2$ fm) and a quark mass $a m_q = 0.0125$ is
shown. This corresponds to a physical mass $m_q \sim 14$ MeV, a bit heavier
than the physical masses of the up and down quarks. On the horizontal axis we
plot $(T/T_{\star})$. We note that the critical temperature for quenched
QCD is $T_\star \sim 260$ MeV \cite{Boyd}, and for two-flavor QCD 
$T_\star \sim 140$ MeV \cite{MILC96}. Unfortunately $\rho(T)$ and $p(T)$ 
for three quarks (u,d,s) with physical
masses is not available yet. For $T/T_{\star} = 4$ energy density and pressure
for quenched QCD are still 10\% resp. 15\% below the ideal gas value. It is
remarkable that $\rho/\rho_{\rm ideal}$ and $p/\rho_{\rm ideal}$ versus
$T/T_{\star}$ is quite similar for quenched QCD and two-flavor QCD.  Moreover,
the temperature dependence of the rescaled pressure for four-flavor QCD
\cite{Karsch97} is quite similar to quenched QCD.

At temperatures below $T_{\star}$ quarks and gluons are confined to hadrons,
mostly pions. At present the hot pion phase is not seen in the two-flavor
lattice QCD, since the pion comes out too heavy ($0.3 <
m_\pi/m_\rho < 0.7$ from \cite{MILC96}, whereas the physical ratio is $0.18$). 
 
\begin{figure}
\begin{center}
\epsfig{figure=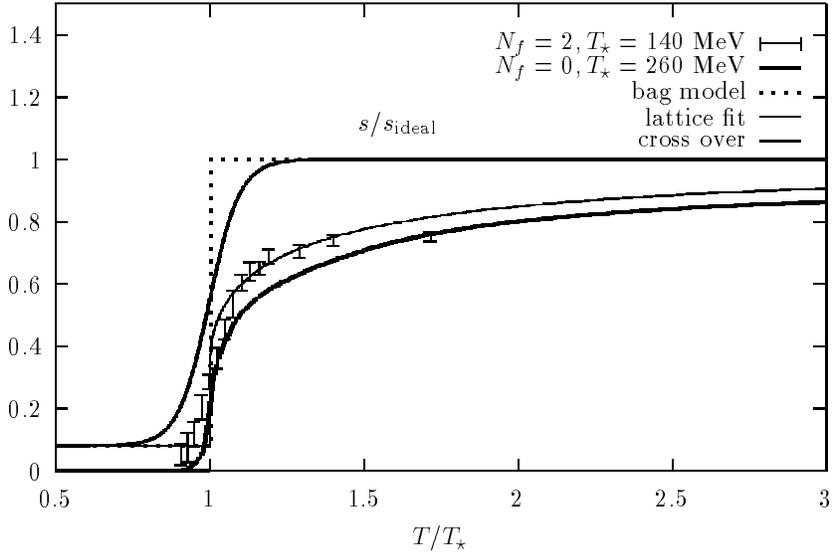,width=0.67\linewidth}
\end{center}
\vspace{-7pt}
\caption{\label{fig2}
The entropy density of hot QCD relative to the entropy density of 
an ideal QGP is shown for the same lattice QCD results as in Fig.~\ref{fig1}.
We also plot the entropy density for our fit to the lattice QCD data,
for the bag model, and for a smooth crossover.}
\end{figure}

The second law of thermodynamics connects pressure and energy density, and
reads for a fluid without chemical potential (no relevant conserved quantum numbers) 
\begin{equation}
  \label{2ndlaw}
  \rho = T {d p \over d T} - p.
\end{equation}
The Maxwell relation for the free energy 
gives the entropy density,
\begin{equation}
   \label{s}
   s ={dp \over dT}.
\end{equation}
{}From homogeneity the free energy density $f(T) = - p(T)$, and $f(T)$
contains the full thermodynamic information. The lattice
QCD results for $s/s^{\rm ideal}_{\rm QGP}$ are shown in  Fig.~\ref{fig2}.

The bag model gives a simple parameterization for $p(T)$ and hence for
$\rho(T), s(T)$. The bag model ansatz represents the short-distance dynamics
by an ideal gas of quarks and gluons and the long-distance confinement effects
by a constant negative contribution to the pressure, the bag constant $B$,
\begin{mathletters}
\label{pressure}
\begin{equation}
\label{pqgp}
p_{\rm QGP}(T) = p_{\rm QGP}^{\rm ideal}(T)  - B \ .
\end{equation}
We include gluons and u,d quarks, which are effectively massless at $T \approx
T_{\star}$, hence $p_{\rm QGP}^{\rm ideal}(T)=(\pi^2/90) g_{\rm QGP}^*
T^4$, where $g^*$ is the effective number of relativistic helicity
states, $g^*= g_{\rm bosons} + 7/8 g_{\rm fermions}$, $g^*_{\rm
QGP} = 37$. The low temperature phase is a hadron gas (HG). We model it as an
ideal gas of massless pions, 
\begin{equation}
\label{phg}
p_{\rm HG} = {\pi^2\over 90} g^*_{\rm HG} T^4,
\end{equation}
because the contribution of pions is small anyway, $g_{\rm HG}^*/g_{\rm
QGP}^*  = 3/37$. 
\end{mathletters}

At the phase transition ($T=T_{\star}$) the quark-gluon phase and the hadron
phase have the same pressure, $p=p_\star$. The stable phase for $T \ne
T_{\star}$ is the one which gives the minimum for the free energy density
$f(T)$, hence a maximum for $p(T)$. The
condition for pressure equilibrium at $T_{\star}$ gives the relation between
$T_{\star}$ and $B$; using (\ref{pqgp}) and (\ref{phg}) we obtain
\begin{equation}
\label{Tstar}
B = {\pi^2\over 90} (g_{\rm QGP}^* - g_{\rm HG}^*) T_\star^4 \ .
\end{equation}
We take $T_\star$ from lattice QCD calculations with two flavors, which give
$T_\star \in (140,160)$ MeV \cite{lattice,MILC96}. This corresponds to a range 
of bag constants $B^{1/4}\in (195,221)$ MeV. This range is in agreement 
with fits to the light-hadron masses, which yield 
$B^{1/4} = 145$ to $245$ MeV (a compilation of various bag model
light-hadron fits can be found in \cite{Farhi}). 
We adopt the value $T_\star = 150$ MeV.

Energy density and entropy density in the bag model follow from
Eqs.~(\ref{2ndlaw}) - (\ref{pqgp}), 
\begin{mathletters}
\begin{eqnarray}
\label{ener}
\rho_{\rm QGP} &=& \rho^{\rm ideal}_{\rm QGP}+ B \\
\label{ent}
s_{\rm QGP} &=& s_{\rm QGP}^{\rm ideal}, 
\end{eqnarray}
\end{mathletters}
with $\rho^{\rm ideal} = (\pi^2/30) g^* T^4$ and $s^{\rm ideal} =
(2\pi^2/45) g^* T^3$ for massless particles. The bag results for
$\rho/\rho^{\rm ideal}_{\rm QGP}$, $p/\rho^{\rm ideal}_{\rm QGP}$,
$s/s^{\rm ideal}_{\rm QGP}$ are shown in Fig.~\ref{fig1} and
Fig.~\ref{fig2}. Note that $s/T^3$ is a simple step function in the bag model.
The QCD transition is first order in the bag model, and the latent heat per
unit volume $l \equiv T_{\star} \Delta s$, is
\begin{equation}
   l = {2 \pi^2 \over 45} \Delta g^{\star} T_{\star}^4 = 4 B,
\end{equation}
where $\Delta g \equiv g_{\rm QGP} - g_{\rm HG}$.

The order of the QCD phase transition and (for a first-order transition) the
magnitude of the latent heat is still a subject of debate. In quenched QCD the
phase transition is of first order \cite{Boyd}. The latent heat was determined
to be $l \approx 1.4 T_\star^4$ \cite{Iwasaki2}. It is useful to take the
ratio $R_L$ of the latent heat to the value $T_{\star} \Delta s^{\rm ideal}$,
where $\Delta s^{\rm ideal}$ is the difference in entropy between an ideal
massless HG and an ideal massless QGP,
\begin{equation}
R_L \equiv {l\over (T_\star \Delta s)^{\rm ideal}} = 
\left\{ \begin{array}{ll}
               1   & \mbox{bag model} \\
               0.2 & \mbox{quenched lattice QCD}
        \end{array}\right. \ .
\end{equation}
For two light quarks it is likely that the transition is a crossover
\cite{MILC96,MILC97}. This is in agreement with theoretical considerations
\cite{Wilczek}, which predict a second-order phase transition in the massless
quark limit. For three light flavors and for the physical relevant case of two
light and one heavy flavor the phase transition is likely to be of first order
\cite{Iwasaki}. This result was obtained using the Wilson quark action,
whereas results with staggered quarks \cite{Brown} indicate a crossover for
the physical quark masses. For four quark flavors the transition is first
order \cite{Karsch97}. Since the latent heat for lattice QCD is known for
quenched QCD only, we decided to use the latent heat ratio $R_L = 0.2$ from
quenched QCD as an indication for the physical case.

We also need an analytic representation of the lattice QCD data. We decided to
fit the entropy density (Fig.~\ref{fig2}) and to derive the other
thermodynamic quantities with Eqs.~(\ref{2ndlaw}) and (\ref{s}). Below
$T_\star$ we consider again an ideal gas of the three pions. We fit the shape
of the lattice QCD data by
\begin{equation}
\label{fit}
{s^{\rm fit}\over (2\pi^2/45) T^3} =
g_{\rm HG}^* + \Delta g^* \theta(T-T_\star)\left[R_L +
\left(1-R_L\right)\left(1 - {T_\star\over T}\right)^\gamma\right]\ . 
\end{equation}
A good fit for our purpose is obtained for $\gamma \in (0.3, 0.4)$. We fix
$\gamma = 1/3$. The fit is plotted in Fig.~\ref{fig2}. 

Finally, we do not neglect the possibility that the QCD transition
could be a smooth crossover. Such a crossover has been modeled by a simple 
interpolation between the ideal values of the entropy density with a tanh.
It was first given in Ref.~\cite{Blaizot} and is used 
in relativistic hydrodynamical simulations of heavy-ion 
collisions \cite{hisim}. The entropy density is written as
\begin{equation}
\label{crossover}
{s^{\rm crossover}\over s^{\rm ideal}_{\rm QGP}} =
1 - \frac12 {\Delta g^*\over g^*_{\rm QGP}} 
\left[1 - \tanh\left(T-T_\star\over \Delta T\right)\right] \ .
\end{equation}
We choose $\Delta T/T_\star=0.1$, see Fig.~\ref{fig2}. This gives the shape of
the initial rise of the lattice data, but it necessarily fails to give the
slow rise at higher temperatures from $s/s^{\rm ideal}_{\rm QGP} \approx 0.6 $
to 1. Comparing this model to the bag model and the lattice fit allows us to
identify phenomena caused by the cosmological QCD transition that are
independent of the order of the transition. The effects for wavelengths close
to the Hubble scale will be shown to be approximately independent of the order
of the transition. 

The early universe at $T$ around $150$ MeV also  contains photons and
relativistic leptons ($e^\pm, \mu^\pm$,
neutrinos) with a pressure
\begin{equation}
p_{{\rm l}\gamma} = {\pi^2\over 90} g_{{\rm l}\gamma}^* T^4 \ .
\end{equation}
where $g_{{\rm l}\gamma}^* =14.25$. 
We use the
subscript minus to denote the value of some quantity at the beginning of the 
transition, i.e.\ when the universe has cooled to $T_\star$ from above,
and the subscript plus to denote the value of the same quantity at the
end of the transition, i.e. when the temperature starts to decrease again.
  
\subsection{Bubble nucleation}

The expansion of the Universe is very slow compared to the strong, 
electro-magnetic, and weak interactions around $T_\star$. To be more explicit,
the rate of the weak interactions is $\Gamma_{\rm w} \sim G_{\rm F}^2 
T_\star^5 \approx 10^{-14}$ GeV, the rate of the electro-magnetic interactions 
is $\Gamma_{\rm em} \sim \alpha^2 T_\star \approx 10^{-5}$ GeV,
and the rate of the strong interactions is $\Gamma_{\rm s} \sim 
\alpha^2_{\rm s}(T_\star) T_\star \approx 10^{-1}$ GeV. These rates have 
to be compared to the Hubble rate $H \sim T_\star^2/m_{\rm Pl} \approx 
10^{-21}$ GeV. Thus, leptons, photons, and the QGP/HG are in thermal and 
chemical equilibrium at cosmological time scales. All components have
the same temperature locally, i.e.~smeared over scales $\lambda \sim  
10^{-7} R_{\rm H}$. At scales $\lambda > 10^{-7} R_{\rm H}$ strongly, weakly, 
and electro-magnetically interacting matter makes up a single perfect
(i.e. dissipationless) radiation fluid. 

In a first-order phase transition the QGP supercools until hadronic bubbles
are formed at $T_{\rm sc} < T_\star$ \cite{Witten,DeGrand,Fuller}. 
Without `dirt' the bubbles nucleate due to thermal fluctuations (homogeneous
nucleation). The probability to nucleate a bubble by a thermal fluctuation
is proportional to $\exp(\Delta S)$, where $\Delta S$ is the change in entropy 
by creating a bubble. The second law relates $\Delta S$ to the minimal
work done in this process, which is the change in the free energy because
the volume and temperature are fixed \cite{LL}. 
The change in free energy of the system by creating a spherical
bubble with radius $R$ is 
\begin{equation}
\Delta F = {4\pi\over 3}(p_{\rm QGP} - p_{\rm HG})R^3 + 4\pi \sigma R^2 \ ,
\end{equation}
where $\sigma$ is the surface tension. Bubbles can grow if they are 
created with radii greater than the critical bubble radius $R_{\rm crit}$. 
Smaller bubbles disappear again, because the free energy gained from the 
bulk of the bubble is more than compensated by the surface 
energy in the bubble wall. $R_{\rm crit}$ is determined from the maximum
value of $\Delta F(R)$, $R_{\rm crit}(T) = 2 \sigma \left[ p_{\rm HG}(T) - p_{\rm QGP}(T)\right]^{-1}$.
At $T_\star$ the critical bubble size diverges, and no bubble can be formed.
Finally, the probability to form a hadronic bubble with critical radius 
per unit volume and unit time is given by
\begin{equation} 
\label{I}
I(T) = I_0 \exp\left( - {\Delta F_{\rm crit}\over T}\right) \ ,
\end{equation}
with $\Delta F_{\rm crit} = 16\pi\sigma^3/[3(p_{\rm HG} - p_{\rm QGP})^2]$.
For small supercooling $\eta \equiv 1 -  T/T_\star \ll 1$ we may evaluate 
$(p_{\rm HG} - p_{\rm QGP})(T)$ by using the second law of thermodynamics,
i.e.\ $p_{\rm HG} - p_{\rm QGP} \approx l \eta$, and thus
\begin{equation}
\Delta F_{\rm crit} \approx {16\pi\over 3} {\sigma^3\over l^2 \eta^2} 
\ .
\end{equation}
For dimensional reasons the prefactor $I_0 \sim T_\star^4$. A more detailed
calculation of $I_0$ within the bag model has been provided in 
\cite{Csernai}. 
It was shown in Ref.~\cite{Christiansen} that the temperature dependence 
of the prefactor $I_0$ can be neglected for the calculation of the 
supercooling temperature $T_{\rm sc}$ in the cosmological QCD
transition. Furthermore a purely numerical prefactor to $T_\star^4$ would be
irrelevant for our purposes as explained after Eq.~(\ref{eta_sc}).
Therefore the probability to form a critical bubble per unit volume and
unit time can be written as 
\begin{equation}
I \approx T_\star^4 \exp\left(-A/\eta^2\right),
\end{equation}
with $A \equiv 16 \pi \sigma^3/(3 l^2 T_\star)$.  

 The surface tension $\sigma$ is a crucial parameter for $T_{\rm sc}$ 
and $R_{\rm crit}$. The absence of surface excitations in hadronic spectra
suggests that $\sigma^{1/3} \ll B^{1/4}$ \cite{Farhi}. In lattice QCD rather
small values are found. Refs.~\cite{Iwasaki2} find that $\sigma \approx 
0.015 T_\star^3$ for quenched lattice QCD. There are no values for unquenched
QCD available yet. However, an upper bound was obtained for the case of
four-flavor lattice QCD in Ref.~\cite{Markum}, i.e., $\sigma < 0.1 T^3_\star$.
Using the results from quenched lattice QCD we find $A = 2.9 \times 10^{-5}$.

After the first bubbles have been nucleated, they grow most probably
by weak deflagration \cite{DeGrand,Kurki-Suonio,Kajantie,Ignatius}.  
The deflagration front (the bubble wall) moves with the velocity $v_{\rm defl}
\ll 1/\sqrt{3}$ \cite{Kajantie92}. The energy that is released from 
the bubbles is distributed into the surrounding QGP by a supersonic shock 
wave and by neutrino radiation. This reheats the QGP to $T_\star$ and 
prohibits further bubble formation. The shock front propagates with a 
speed slightly above the sound speed. The amplitude of the shock is very small 
\cite{Kurki-Suonio}. Neutrinos have a mean free path of $10^{-7} T_{\rm H}$ at
$T_\star$. If they do most of the heat transport, it goes with
$v_{\rm heat}= {\cal O}(c)$. A detailed calculation of the supercooling
temperature in the cosmological QCD transition was given in
Refs.~\cite{Fuller,Ignatius}. These calculations assume small
supercooling. The suppression of bubble nucleation due to already existing
bubbles is neglected. 

 The supercooling temperature fraction $\eta_{\rm sc}$ turns out to be about
the same for the schematic case of one single bubble nucleated per Hubble
volume per Hubble time,
\begin{equation}
\label{eta_sc}
  \eta_{\rm sc} \approx \left[\frac{A}{4 \ln(T_\star/H_\star)} \right]^{1/2}
\approx 4 \times 10^{-4}  \vspace{2cm}
\end{equation} 
or the realistic case obtained below of one bubble nucleated per ${\rm cm}^3$
per $10^{-6}$ of a Hubble time, which needs a supercooling $20 \%$
larger. This small increase of $20\%$ in $\eta_{\rm sc}$ makes the bubble
nucleation rate larger by a factor $10^{24}$. 

The time needed for the supercooling is given by $\Delta t_{\rm sc}/t_{\rm H}
= \eta_{\rm sc}/(3 c_s^2)= {\cal O}(10^{-3})$. The critical size of the
bubbles created at the supercooling temperature is
\begin{equation}
R_{\rm crit}(\eta_{\rm sc}) \approx {2\sigma\over l \eta_{\rm sc}} \approx
               30 \mbox{ fm}  \vspace{2cm}.
\end{equation}

Bubbles present at a given time have been nucleated typically during the
preceding time interval $\Delta t_{\rm nucl} \equiv  I/({\rm d}I/{\rm d} t) = 
{\cal O}(10^{-6})$. Using the relation between time and supercooling
$\eta$, ${\rm d}\eta/{\rm d}t = 3 c_s^2 /t_{\rm H}$, we find $\Delta t_{\rm
nucl}/t_{\rm H}= \eta^3/(6 A c_s^2)$. During this time interval each
bubble has distributed released latent heat over a distance $\approx \Delta
t_{\rm nucl} v_{\rm heat}$. This distance has a weak dependence on the precise
value of $\eta_{\rm sc}$, but the bubble nucleation rate increases exceedingly
strongly with $\eta$ until one bubble per volume $\sim (\Delta t_{\rm nucl}
v_{\rm heat})^3$ is nucleated. Therefore the mean bubble separation is
\begin{equation}
\label{dnucl}
d_{\rm nucl} \approx v_{\rm heat} \Delta t_{\rm nucl} \approx \frac{v_{\rm heat}}{3 c_s^2}  {\eta_{\rm sc}^3\over
A}  R_{\rm H}= {\cal O}(10^{-6} R_{\rm H})= {\cal O}(1 {\rm cm}), 
\end{equation}
where we used $v_{\rm heat} = {\cal O}(c), c_s = {\cal O}(c)$, which gives a
typical value for the nucleation distance. In \cite{Christiansen} the mean
bubble nucleation distance due to thermal fluctuations (homogeneous
nucleation) is calculated to be less than $2$ cm, whereas for inhomogeneous
nucleation (the first bubbles form at impurities like topological defects or
primordial black holes) it might be a few m. 

The above estimate of bubble separation applies if the limiting factor for
quenching is the distribution of released latent heat by means of sound waves
and by neutrino free streaming. On the other hand the limiting factor could be
given by the rate of release of latent heat, i.e. by the bubble wall velocity
$v_{\rm defl}$. Since the period of supercooling lasts about $1 \% $ of the
time needed for completing the entire first-order phase transition, $1 \% $ of
the QGP must be converted to HG in the process of sudden reheating to
$T_{\star}$, hence the bubble radius at quenching must reach $0.2$ of the
bubble separation, $R_{\rm bubble}\approx 0.2 d_{\rm nucl}$.  With $R_{\rm
bubble} \approx v_{\rm defl} \Delta t_{\rm nucl}$ and using the above relation
$d_{\rm nucl} \approx v_{\rm heat} \Delta t_{\rm nucl}$, we require $v_{\rm
defl} \ge 0.2 v_{\rm heat}$ for consistency. If $v_{\rm defl}$ is smaller than
this, the limiting factor for quenching is the rate of 
release of latent heat by bubble growth, and the bubble separation is  
\begin{equation}
d_{\rm nucl} \approx v_{\rm defl} \Delta t_{\rm nucl} \approx \frac{v_{\rm defl}}{3 c_s^2}  {\eta_{\rm sc}^3\over
A} R_{\rm H}, 
\end{equation}
i.e. the bubble separation will be smaller than the estimate
Eq.~(\ref{dnucl}). 

To summarize, the scales on which non-equilibrium phenomena occur are 
given by the mean bubble separation, which is about $10^{-6} R_{\rm H}$. 
The entropy production is tiny, i.e. $\Delta S/ S \sim 10^{-6}$, since the 
supercooling is small $\sim 10^{-3}$. After supercooling, which lasts $10^{-3}
t_{\rm H}$, the universe reheats in $\Delta t_{\rm nucl} \approx 10^{-6}
t_{\rm H}$. After reheating the thermodynamic variables follow their
equilibrium values and bubbles grow due to the expansion of the Universe only.
This is in striking contrast to bubble nucleation at the GUT transition, where
the particle interaction rate is of the order of the Hubble rate and the
bubble nucleation rate per unit volume is very much smaller than $H^4$.

\subsection{Adiabatic phase conversion}
\begin{figure}
\begin{center}
\epsfig{figure=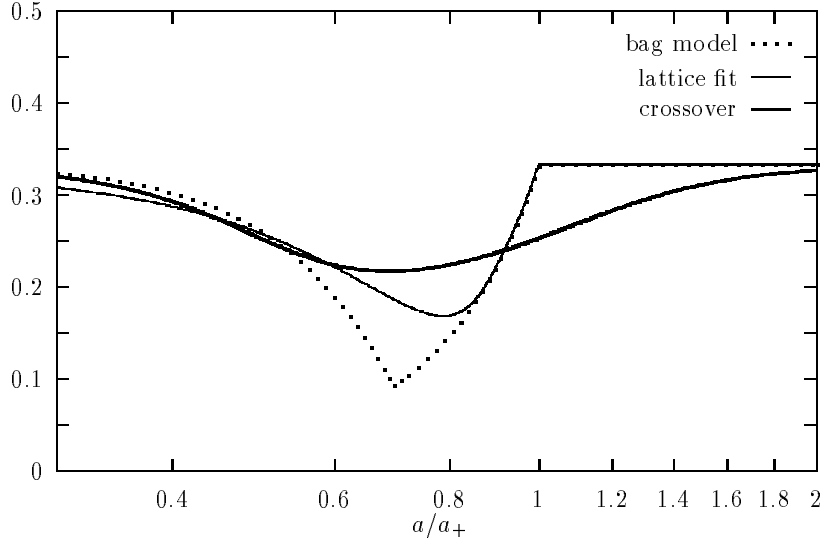,width=0.67\linewidth}
\end{center}
\caption{\label{fig3}
The behavior of $w=p/\rho$ during the QCD transition as a function of the
scale factor $a$. Although the pressure is reduced, it stays positive 
throughout the transition.}
\end{figure}

We now discuss the evolution of the temperature and other
thermodynamic variables as a function of the scale factor $a$. The evolution
of $w\equiv p/\rho$ for the bag model, our lattice QCD fit, and for the smooth
crossover (\ref{crossover}) is shown in Fig.~\ref{fig3}.

Entropy is conserved, apart from the very short stage of reheating 
($\sim 10^{-6} t_{\rm H}$) after the first bubbles have been nucleated.
This allows us to calculate $T=T(a)$ from ${\rm d}\left[s(T) a^3\right] = 0$,
i.e. 
\begin{equation}
\label{Ta}
{{\rm d}T\over{\rm d}\ln a} = - {3 s\over {\rm d} s/{\rm d} T} \ ,
\end{equation}
except for $T = T_\star$ in the case of a first-order phase transition. 
In the bag model $s \sim T^3$ and therefore $T \propto 1/a$ for $T\neq
T_\star$. In the case of the lattice fit and the crossover the entropy is
given by Eq.~(\ref{fit}) resp. Eq.~(\ref{crossover}) plus the contribution of
photons and leptons. Pressure as a function of the scale factor is obtained
from $s = dp/dT$, using Eq.~(\ref{Ta}). The energy density, hence $w$, follows
from Eq.~(\ref{2ndlaw}). 

While the QGP and HG  coexist in a first-order QCD transition the expansion 
factor is determined by entropy conservation,
\begin{equation}
{a_+ \over a_-} = \left({s_-\over s_+}\right)^\frac13  \ ,
\end{equation}
where the index $-(+)$ denotes the value of a quantity at the beginning
(end) of the coexistence epoch. In the bag model the Universe expands by
a factor $a_+/a_- \approx 1.4$ until all QGP has been converted into the HG,
whereas for our lattice QCD fit (\ref{fit}) the Universe expands by a
factor $a_+/a_- \approx 1.1$. The growth of the scale factor is related to a
lapse in cosmic time by ${\rm d}\ln a = H {\rm d} t$. In terms of Hubble time
the transition lasts $0.3 t_{\rm H}$ for the bag model and $0.1 t_{\rm H}$ for
 our lattice QCD fit. 

During a first-order QCD transition, i.e.~$T = T_\star$, the pressure
$p(T_\star) \equiv p_\star$ is constant. For any first-order QCD phase
transition $\rho(a)$ is obtained from the first law
of thermodynamics $d\rho = -3(\rho + p_\star)da/a$. The result
for $a_- \le a \le a_+$ is:
\begin{eqnarray}
\rho(a) &=& (\rho(a_-) + p_\star)
\left({a_-\over a}\right)^3 - p_\star \ , \\
w(a) &=& \left[\left({\rho(a_-)\over p_\star} + 1 \right)
\left({a_-\over a}\right)^3 - 1\right]^{-1} \ ,
\end{eqnarray}
where $w\equiv p/\rho$.
After the end of the phase transition we have made the approximation of
noninteracting and massless pions, leptons and photons, $p=\rho/3$, and
therefore $p_\star = \rho(a_+)/3$.

\subsection{The sound speed}

\begin{figure}
\begin{center}
\epsfig{figure=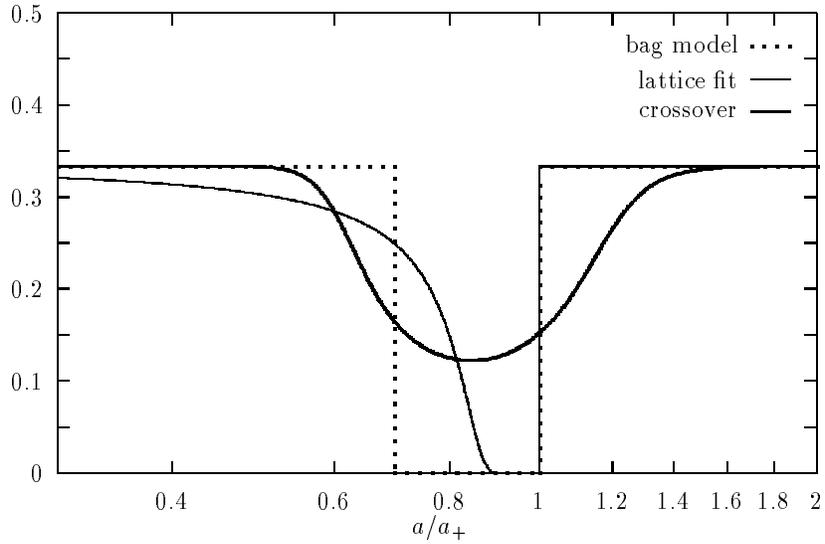,width=0.67\linewidth}
\end{center}
\caption{\label{fig4}
The behavior of the sound speed $c_s^2 = (\partial p/
\partial\rho)_s$ during the QCD transition as a function of $a$.
For a first-order transition (lattice fit and bag model) the sound speed
vanishes.}
\end{figure}

As explained in the introduction, the isentropic sound speed (for wavelengths
$\lambda$ much larger than the bubble separation), given by 
\begin{equation}
\label{cs2}
c_s^2 = \left({\partial p\over \partial \rho}\right)_s =
\left({dp/da\over d\rho/da}\right)_s \ ,
\end{equation}
must be zero during a first-order phase transition for a fluid with negligible
chemical potential (i.e. no relevant conserved quantum number). This behavior
is shown in Fig.~\ref{fig4} for the lattice QCD fit and the bag model. In the
bag model $c_s^2 =1/3$ before and after the transition. For a crossover the sound speed does not drop to zero.

A strong decrease in the sound speed is observed in lattice QCD 
for $N_f = 0$ \cite{Boyd} and for $N_f = 2$ \cite{MILC96}. From our lattice 
QCD fit we can calculate the sound speed for $T > T_\star$. From
Eq.~(\ref{2ndlaw}) and Eq.~(\ref{s}) the sound speed is
given by $c_s^2 = ({\rm d}\ln s/{\rm d}\ln T)^{-1}$. Inserting (\ref{fit}), we
obtain 
\begin{equation}
c_s^2 \propto \left(1 - {T_\star\over T}\right)^{1-\gamma} \ .
\end{equation}
For an analytic discussion of the behavior of acoustic modes during the
phase transition it is important to know $c_s^2$ as a function of 
conformal time $\eta \equiv \int a^{-1}(t) {\rm d}t$. 
In the radiation dominated regime $a \propto \eta$.
Up to small corrections, this holds true during the transition. 
Let us denote the value of conformal time at which the sound speed vanishes
by $\eta_-$. From Eq.~(\ref{Ta}) we find $(\eta_-  - \eta)/\eta_- \propto 
(1 - T_\star/T)^\gamma$, which implies
\begin{equation}
\label{cseta-}
c_s^2 = C^2 \left[ (\eta_-  - \eta)/\eta_- \right]^{1 -\gamma\over\gamma} \ , 
\end{equation} 
with $C^2 = \{ 3^{1-\gamma} [g^*_+ + \Delta g^* R_L]/
[\Delta g^* (1 - R_L)]\}^{1/\gamma}/\gamma$.
For $\gamma = 1/3$ the sound speed goes to zero linearly in conformal
time. 

We now discuss why the isentropic condition applies during the part of the
phase transition after the initial supercooling, bubble nucleation, and sudden
reheating to $T_{\star}$. During the second part of the transition, which
takes about $99\%$ of the transition time, the fluid is extremely close to
thermal equilibrium, because the time to reach equilibrium is very much
shorter than a Hubble time, i.e. the fluid makes a reversible
transformation. This can be seen as follows. Across the bubble walls local
pressure equilibrium is established immediately, $p_{\rm QGP} = p_{\rm HG}$
locally. Local temperature equilibrium, $T_{\rm QGP} = T_{\rm HG}$ is
established by neutrinos, which have a mean free path of $10^{-7} R_{\rm H}$,
enormously larger than the bubble wall thickness, and a collision time much
shorter than the Hubble time. This local pressure and temperature equilibrium
can only be satisfied if $p=p_{\star}$ and $T=T_{\star}$ at the bubble
walls. Over distance scales of the order of the bubble separation ($\sim 1$
cm) pressure (and therefore also temperature) is equalized with the velocity
of sound, and thereby the released latent heat is distributed. This pressure
equalization is very fast compared to the Hubble expansion velocity $v_{\rm H}
\approx 10^{-6} c$ at the $1$ cm scale. When analyzing cosmological
perturbations we shall consider wavelengths $\lambda > 10^{-4} R_{\rm H}$, for
which neutrinos are tightly coupled, $\Gamma_{\nu}/k \gg 1$, see Sec.~IV. For
these wavelengths the radiation fluid behaves as a perfect
(i.e. dissipationless) fluid, entropy in a comoving volume is conserved, one
has a reversible process. On the other hand, below the neutrino diffusion
scale, $\lambda < 10^{-4} R_H$, acoustic oscillations are damped away before
the QCD transition.

We have been criticized in \cite{jap} for the use of the isentropic
condition. In \cite{jap} it was claimed that the isentropic condition leads to
an infinite sound speed in the bag model. This claim was wrong, because it was
based on non-relativistic hydrodynamical equations which neglect the pressure
contribution to the momentum density and neglect that sound waves in a
radiation fluid move with relativistic speed. In a revised version of their
work \cite{jap2} they correct these mistakes. However they still maintain that
the isentropic condition does not hold, but that instead the volume fraction
of the two phases is frozen. They do not give any argument why such a freezing
should occur, and their freezing assumption contradicts the rapid
equilibration of pressure and temperature between the two phases explained
in the previous paragraph. Their freezing assumption would also entail
supercooling from the Hubble expansion in the presence of a finite volume
fraction of bubbles. This is in dramatic conflict with the 
extremely rapid reheating to $T_{\star}$ (within $10^{-6} t_{\rm H}$) once
enough bubbles ($\sim 1\%$ of volume) have been formed as discussed in
Sec.~IIB. 

\section{Peaks and dips in the density spectrum}

\subsection{Evolution equations for cosmological perturbations
in uniform expansion gauge}

The evolution of cosmological perturbations at the QCD transition is analyzed
for perfect fluids. We linearize the Einstein equations, e.g. $\rho({\bf
x},t)=:\rho_0(t)+\delta \rho({\bf x},t)$.  It is an excellent approximation at
early times to take the Friedmann-Robertson-Walker metric flat, $ds^2 = -
dt^2+a(t)^2 d {\bf x}\cdot d {\bf x}$. The evolution of the background
quantities is given by the Friedmann equations,  
\begin{eqnarray}
\label{Friedmann}
 H^2     &=&\frac{8 \pi G}{3} \rho_0 \\\nonumber
 \dot{H} &=& -4 \pi G (\rho_0+p_0),
\end{eqnarray}
where the subscript 0 denotes background quantities. 

The perturbed metric is decomposed in a scalar, a vector, and a tensor part. We
are only interested in density perturbations, which means the scalar
(longitudinal) sector. For a time-orthogonal foliation of space-time
the perturbed metric can be written as 
\begin{equation}
\label{metric}
ds^2 = -\left( 1+ 2 \alpha \right) d t^2 + a(t)^2 \left[ \delta_{i j}(1+2
\varphi) +2 \partial_i\partial_j \gamma \right] dx^i dx^j,
\end{equation}
with $\partial_i \equiv {\partial \over \partial x^i}$. We follow Bardeen 1989
\cite{Bardeen} and focus on the geometric properties of the constant-t
hypersurfaces (slices) $\Sigma$. The perturbation of the lapse function
between the $\Sigma$'s is given by $\alpha ({\bf x},t)$. The intrinsic
curvature of $\Sigma$ is generated by $\varphi ({\bf x},t)$,
\begin{equation}
 ^{(3)}R[\Sigma] = -4 \bigtriangleup \varphi.
 \end{equation}
The extrinsic curvature $K^i_{\, \,j}$ of $\Sigma$ has 3 terms, the
unperturbed term, the perturbation of the isotropic part, $\delta[\mbox{tr}
K^i_{\,\,j}(\Sigma)]\equiv\kappa$, and the anisotropic part ($\equiv$ shear of
normals on $\Sigma$), which is generated by $\chi({\bf x},t)$,
\begin{equation}
 K^i_{\, \,j}[\Sigma] = -H \delta^i_{\,\,j}+\frac13 \kappa  \delta^i_{\,\,j}
-(\partial^i \partial_j - \frac13 \bigtriangleup  \delta^i_{\,\,j})\chi, 
\end{equation}
where $\partial^i=a^{-2} \delta^{ij}\partial_i$.
These geometric properties of $\Sigma$ are computed from the metric
perturbations as follows, $\chi = a^2 \dot{\gamma}$, $\kappa =
-3\left(\dot{\varphi}-H \alpha\right) - \bigtriangleup \chi$. Vanishing shear
of $\underline{n}(\Sigma)$ means $\chi=0$ and vice versa, but note that $\chi$
(resp. $\gamma$) contributes both to the anisotropic part and to the isotropic
part of $K^i_{\, \,j}$.

We choose a slicing of space-time with unperturbed mean extrinsic curvature,
$\delta[\mbox{tr} K^i_{\,\,j}(\Sigma)]\equiv\kappa=0$. This implies that our
fundamental observers, which are defined to be at rest on the slice $\Sigma$,
$\underline{u}(\mbox{obs})=\underline{n}(\Sigma)$, have relative velocities,
which in the mean over all directions follow an unperturbed Hubble flow. If
the coordinate choice (gauge choice) is such that the time coordinate $t$ is
constant on the slices $\Sigma$, the gauge is fixed to be the Uniform
Expansion Gauge (UEG) $\equiv$ Uniform Hubble Gauge (UHG) \cite{Bardeen}.

In the literature the most common gauges are the synchronous gauge, where the
fundamental observers are moving inertially (i.e. $\alpha \equiv 0$), the
longitudinal gauge ($\equiv$ zero shear gauge, $\chi\equiv0$), where the
fundamental observers expand isotropically, and the comoving gauge, where the
fundamental observers sit in the fluid rest frame, $\underline{u}(\mbox{obs})
= \underline{u}(\mbox{matter})$. The uniform expansion gauge (UEG) is free of
spurious gauge modes (in contrast to the synchronous gauge), it is
nonsingular in the superhorizon limit and therefore easy to handle in numerical
calculations (in contrast to zero-shear variables, as pointed out by Bardeen
\cite{Bardeen}), and it has the two basic fluid variables $\delta \rho$ and
$v_{\rm peculiar}$ (in contrast to the comoving gauge).

As fundamental evolution equations in UEG for perfect fluids we take the
energy-momentum conservation $\nabla_{\mu} T^{\mu}_{\ \nu} = 0$, i.e., the
continuity equation and  (in the longitudinal sector) the 3-divergence of the
Euler equation of general relativity. Each fluid separately must fulfill
\begin{eqnarray}
\label{C}
\partial_t \epsilon &=& - 3 H(\epsilon + \pi) - \bigtriangleup \psi -
3 H (\rho_0 + p_0)\alpha \\
\label{E}
\partial_t \psi &=& - 3 H \psi - \pi - (\rho_0 + 
p_0)\alpha \ ,
\end{eqnarray}
where $\epsilon \equiv \delta\!\rho = \delta T^{0 }_{\; \;0}$, $\pi
\equiv \delta\! p = 1/3 \delta[\mbox{tr} T_{\ j}^{i}]$, $\nabla_i \psi
\equiv \delta T^{0 }_{\; \; i}$ for each fluid separately. Note that
$\nabla_i \psi \equiv S_i=$ momentum density (Poynting vector). 
The system of dynamical equations is closed by Einstein's 
$R^0_{\ 0}$-equation,  
\begin{equation}
\label{R}
(\bigtriangleup + 3 \dot{H})\alpha = 4\pi G (\epsilon + 3 \pi) \ ,
\end{equation}
where the sum over all fluids is taken on the right-hand side. In addition we
need $(p/\rho)$ and $(\pi/\epsilon)_s=c^2_s$, Figs.~\ref{fig3} and \ref{fig4},
for adiabatic perturbations. Equations (\ref{C}) -- (\ref{R}) define our
general relativistic Cauchy problem (linear perturbations, longitudinal
sector, perfect fluids). The initial data $(\epsilon, \psi)$ can be chosen
freely on $\Sigma_{\rm UE}$.

Einstein's $R^0_{\ 0}$-equation in linear perturbation theory in UEG is 
an elliptic equation. Its Green function is
a Yukawa potential with exponential cutoff at the Hubble radius, $\alpha \sim
r^{-1} \exp(-\mu r)$, where $\mu ^2 = -3 \dot{H}$, instead of Newton's
$1/r$-potential in the Jeans equations. $R^0_{\ 0}$ gives the relative
gravitational acceleration (geodesic deviation), averaged over all directions,
of two test particles, which are moving nonrelativistically in the
coordinate system at hand. In UEG $\delta R^0_{\ 0}$ is determined by
$\alpha$, the perturbation of the lapse. In Newtonian physics the relative
acceleration averaged over directions is given by $\Delta \Phi_{\rm Newton}$,
i.e. $R^0_{\ 0}$ and $\Delta \Phi_{\rm Newton}$ answer the same physics
question. In UEG $\alpha$ plays the role of the gravitational potential.

Our basic equations have exactly the same structure as the Jeans equations:
two first-order time-evolution equations for each fluid,
Eqs.~(\ref{C}) and (\ref{E}), supplemented by one elliptic equation,
Eq.~(\ref{R}). The UEG is the only gauge with
this structure and with Einstein's equation appearing only in the elliptic 
equation. In the subhorizon limit (drop the last term in (\ref{C}) and the
$\dot{H}$-term in (\ref{R})) and with $p \ll \rho$ these equations
automatically reduce to the Jeans equations. The UEG is
singled out for two reasons: First, the relative velocity of the fundamental
observers in the mean over all directions is an unperturbed Hubble flow both
in the UEG of general relativity and in the Jeans
analysis. Second, only the mean over all directions appears in the continuity
equation, in the 3-divergence of the Euler equation, and in the  $R^0_{\
0}$-equation.  

Out of the geometric quantities $(\kappa,\chi,\varphi,\alpha)$ only the lapse
$\alpha$ appears in the equations needed to solve the dynamics in UEG. The
spatial curvature ($\varphi$) and the shear of the fundamental observers
($\chi$) are automatically absent from the above equations for perfect fluids,
since these geometrical properties of $\Sigma$ are irrelevant for the
dynamical question at hand. ($\varphi,\chi$) are not needed at all to solve
the dynamics. One can compute $\varphi$ and $\chi$ at any time
from the energy and the momentum constraints, i.e. from the $G^0_{\ 0}$
equation and the longitudinal contribution to the $G^0_{\ i}$ equation,
\begin{eqnarray}
\bigtriangleup \varphi &=& - 4\pi G \epsilon \label{energyconstr}\\
\bigtriangleup \chi    &=& - 12\pi G \psi \ .
\end{eqnarray}  
This sequence, solving the dynamics before using the $G^0_{\ 0}$ and $G^0_{\
i}$ constraint equations, is different from usual solution strategies in
linear perturbation theory. 

Since the evolution equations are linear, and the background is spatially
homogeneous, each spatial Fourier mode with comoving wavelength $k:=k^{\rm
phys} a$ evolves independently. It is convenient to rewrite equations
(\ref{C}) -- (\ref{R}) in the dimensionless variables $\delta \equiv
\epsilon/\rho$ (density contrast), $\hat{\psi} \equiv k_{\rm phys} 
\psi/\rho$. The variable $\hat{\psi}$ is related to the fluid velocity
(peculiar velocity) $v_{\rm pec} = \frac{\rho}{\rho+p} |\hat{\psi}|$ and has
the same order of magnitude as $\delta$ on subhorizon scales, $\lambda^{\rm
phys} \ll R_{\rm H}=H^{-1}$. In UEG the system of evolution equations,
written in terms of conformal time $( )^\prime \equiv \partial_{\eta}
\equiv a \partial_t$, now reads
\begin{eqnarray}
\label{C1}
\frac{1}{\mathcal{H}}\delta^{\prime}+3(c^2_s-w)\delta 
&=&\frac{k}{\mathcal{H}}\hat{\psi} -3(1+w)\alpha\\
\label{E1}
\frac{1}{\mathcal{H}}\hat{\psi}^{ \prime}+(1-3w)\hat{\psi} &=& 
-c_s^2\frac{k}{\mathcal{H}}\delta- (1+w)\frac{k}{\mathcal{H}}\alpha \\
\label{R1}
\left[\left(\frac{k}{\mathcal H}\right)^2 + \frac{9}{2}(1+w)\right]\alpha &=& 
     -\frac{3}{2}(1+3 c_s^2) \delta,
\end{eqnarray}
with ${\mathcal H} \equiv (da/d\eta)/ a$ and $w = p/\rho$. The Friedmann
equation and the continuity equation for the background  were used. The
continuity and Euler equation refer to each fluid separately. In the $R^0_{\
0}$-equation $w=(\sum_f p^{(f)})/(\sum_f \rho^{(f)})$, $\delta = (\sum_f
\epsilon^{(f)})/(\sum_f \rho^{(f)})$ and $c_s^2 = (\sum_f \pi^{(f)})/(\sum_f
\epsilon^{(f)})$. 

In our numerical analysis we have used the exact general relativistic
equations  (\ref{C1}) -- (\ref{R1}), but it is instructive to look at the
superhorizon $\lambda^{\rm phys} \gg H^{-1}$ and at the subhorizon
$\lambda^{\rm phys} \ll H^{-1}$ limit of equations (\ref{C1}) -- (\ref{R1})
for general pressure and sound speed.

For superhorizon perturbations with arbitrary equation of state (e.g. QCD
transition), we only have to keep the lowest order in $k/{\mathcal H}$ in the
evolution equations. The Equations (\ref{C1}) -- (\ref{R1}) for a one fluid
model simplify to   
\begin{eqnarray}
\frac{1}{\mathcal H}\delta' -(1+3w)\delta &=&\frac{k}{\mathcal H}\hat{\psi} \\
\frac{1}{\mathcal H}\hat{\psi}'+(1-3w)\hat{\psi} &=&
\frac{1}{3}\frac{k}{\mathcal H}\delta,
\end{eqnarray}
where the superhorizon limit of the Poisson equation $(1+w)\alpha =
-\frac{1}{3}(1+3 c_s^2)\delta$ was inserted. Note that the sound speed drops
out in the superhorizon equations. For the growing mode the terms on the left
hand side of the Euler equation add and $\left[\hat{\psi}/\delta\right]_{\rm
grow} \propto k/{\mathcal H}$. Therefore, the right hand side of the continuity
equation is negligible and the solution for the growing mode of the
density contrast is 
\begin{equation}
\delta \propto \left(k/{\mathcal H}\right)^2= \left(k_{\rm phys}/H\right)^2 
   \label{super}
\end{equation} 
for superhorizon perturbations in the uniform expansion gauge.
This result is independent of the thermodynamic behavior. It says
that the evolution of the growing superhorizon mode, if expressed in terms of
${\mathcal H}$ (as opposed to expressing it in terms of $t$ or $a$), is
unaffected by a phase transition. Therefore the QCD-transition cannot affect
superhorizon scales, and the spectrum of density perturbations stays flat for
$\lambda^{\rm phys} \gg H^{-1}$. This result can also be formulated
differently: using the energy-constraint (\ref{energyconstr}) we find $\varphi
= const$ for the dominant mode on superhorizon scales in UEG. This is the
'conservation law' \cite{BST} for the quantity $\zeta \equiv \varphi +
\frac{1}{3(1+w)}\delta$ for $w \ne -1$, since $\zeta \approx \varphi$ in the
superhorizon limit.   
        
For subhorizon analysis, $\lambda^{\rm phys} \ll H^{-1},$ with arbitrary
equation of state, we can drop $\dot{H}$ in the general relativistic Poisson
equation (\ref{R}) since $|\dot{H}| = O(H^2) \ll k^2_{\rm phys}$. In the
continuity equation (\ref{C}) the time dilation term (last term) can be
omitted since it follows from the Poisson equation that $\alpha
=O(H^2/k^2_{\rm phys}) \delta \ll \delta$. In the subhorizon limit
Eqs.~(\ref{C1}) -- (\ref{R1}) simplify to 
\begin{eqnarray}
\label{C2}
\frac{1}{\mathcal H}\delta'+3(c_s^2-w)\delta &=&\frac{k}{\mathcal H}\hat{\psi} \\
\label{E2}
\frac{1}{\mathcal H}\hat{\psi}'+(1-3w)\hat{\psi} &=&  -c_s^2\frac{k}{\mathcal H}\delta -
(1+w)\frac{k}{\mathcal H}\alpha \\ 
\label{R2}
\left(\frac{k}{\mathcal H}\right)^2 \alpha &=& -\frac{3}{2}(1+3 c_s^2)\delta. \ 
\end{eqnarray}
 In the case of the QCD transition, the transition time is short compared to
the Hubble time $(t_+ - t_-) < t_{\rm H}=H^{-1}$ and therefore self-gravity of
the dominant radiation fluid can 
be neglected in the Euler equation (\ref{E2}) for this component. On the other
hand, CDM moves in the external gravitational potential provided by the
radiation fluid, and is coupled to the radiation fluid via the
gravity term in the Euler equation. Therefore we have to keep the gravity term
in the CDM Euler equation. 

For a purely radiation dominated universe $w=c_s^2=\frac13$, the solution 
for the density contrast $\delta$ and the peculiar velocity $\hat{\psi}$ can 
be written in terms of spherical Bessel and Neumann functions \cite{Hwang},
\begin{eqnarray}
\delta &=& x f(x) + \frac{3}{x^2+6} x^2 \frac{d}{dx}f(x)\\
\hat{\psi} &=& \frac{1}{x^2+6} x^3 \frac{d}{dx}f(x),
\end{eqnarray}
where  $x \equiv k_{\rm phys}/H = k/{\mathcal H}$. For a
radiation dominated universe $x=k\eta$, $f(x)\equiv \left[A
j_1(x/\sqrt{3})+B n_1(x/\sqrt{3})\right]/\sqrt{3}$. The normalization is
chosen in such a way that the subhorizon modes, $\lambda^{\rm phys} \ll
H^{-1}$, of  $\delta$ and $\sqrt{3}\hat{\psi}$ are acoustic oscillations with
constant amplitude $\sqrt{A^2+B^2}$. We will fix our initial conditions to be
the growing mode in the radiation dominated superhorizon regime, 
\begin{eqnarray}
\label{initialcond1}
\delta &=& A \frac16 x^2\\
\label{initialcond2}
\hat{\psi} &=& A \frac{1}{54} x^3.
\label{initialcond}
\end{eqnarray}

\subsection{Numerical results}

The transfer functions for the radiation fluid and the cold dark matter (CDM)
are calculated by integrating the exact general relativistic set of equations
(\ref{C1})--(\ref{R1}) throughout the QCD transition. The initial
conditions are fixed at $T=100$ GeV, where all scales under considerations
were far above the horizon. We fix the initial amplitude of the radiation
fluid, $A_{\rm in}$, by the growing mode of an exactly radiation dominated
solution, Eqs.~(\ref{initialcond1}) and (\ref{initialcond2}). The normalization
constant is given by the COBE normalized \cite{COBE} Harrison-Zel'dovich
spectrum \cite{HZ}, but it drops out in the transfer functions. 
The initial conditions for CDM are obtained assuming adiabatic
perturbations, i.e. the entropy per cold particle is unperturbed $\delta
\left(s^{\rm RAD}/n^{\rm CDM}\right) = 0$. Since $\rho^{\rm CDM}= m n^{\rm
CDM}\propto 1/V_{\rm comoving}$ and $\rho_{\rm RAD} \propto \left(1/V_{\rm
comoving}\right)^{4/3}$, the adiabatic initial conditions for $\delta^{\rm CDM}$ can
be written   
\begin{equation} 
\delta^{\rm CDM}=\frac34 \delta^{\rm RAD}.
\end{equation} 
The initial fluid velocities are equal. Since we are in the linear regime, each
spatial Fourier mode evolves independently. 

For a first-order phase transition it is convenient to use the volume of
a comoving box $\propto a^3$ instead of time as the evolution variable. 
Thus, the derivatives in Eqs.~(\ref{C1}) and (\ref{E1})
have to be replaced, ${\mathcal H}^{-1} (d/d \eta) = a (d/d a)$. In sections
II.C. and II.D. we have given $w$ and $c_s^2$ in terms of the scale factor. In
the case of a crossover we use the temperature $T$ as the independent 
evolution variable, since temperature is decreasing strictly monotonically in
time. The derivatives in Eqs.~(\ref{C1}) and (\ref{E1}) are replaced by
${\mathcal H}^{-1} (d/d \eta)= -3 c_s^2 T (d/d T)$. We obtain $p(T)$ by
integrating the entropy density $s(T)$ , Eq.~(\ref{s}).

The transfer functions are evaluated by evolving each mode from 
the initial temperature $T_{\rm in}=100$ GeV to the final temperature
$T_{\rm f}=T_{\star}/10 = 15$ MeV, i.e. until the peculiar
velocity of CDM has redshifted away. At $T_{\rm f}$ the universe is exactly
radiation dominated. The amplitude of the acoustic oscillations is 
$A^{\rm RAD} \equiv (\delta_{\rm RAD}^2 +3 \hat{\psi}_{\rm RAD}^2)^{1/2}$. 
The final amplitude of CDM is $A^{\rm CDM} = |\delta^{\rm CDM}|$. The transfer 
functions for the radiation and the CDM fluid are defined by 
\begin{eqnarray} 
  {\mathcal T}^{\rm RAD}(k,{\rm in}\rightarrow {\rm f}) 
  \equiv \frac{A^{\rm RAD}(k)|_{T_{\rm f}}}{A_{\rm in}}\\
  {\mathcal T}^{\rm CDM}(k,{\rm in}\rightarrow {\rm f}) 
  \equiv \frac{A^{\rm CDM}(k)|_{T_{\rm f}}}{A_{\rm in}} \ ,
\end{eqnarray}
where $A_{\rm in}=A_{\rm in}^{\rm RAD}$.

\begin{figure}
\begin{center}
\epsfig{figure=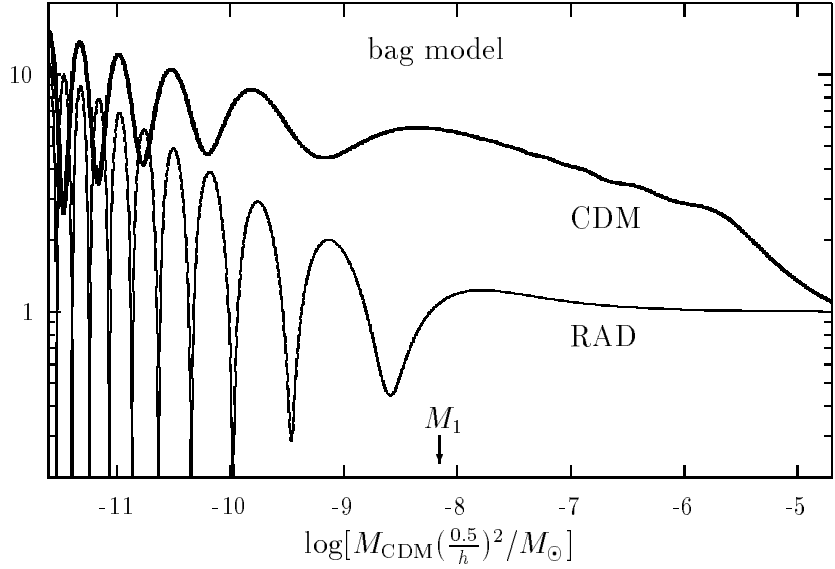,width=0.67\linewidth}
\end{center}
\caption{\label{fig5} 
The modifications of the density contrast for kinetically decoupled CDM (like
axions or primordial black holes) $A^{\rm CDM} \equiv | \delta^{\rm CDM}|
(T_\star /10)$ and of the radiation fluid amplitude $A^{\rm RAD} \equiv
(\delta_{\rm RAD}^2 + 3 \hat{\psi}_{\rm RAD}^2)^{1/2}$  due to the QCD
transition in the bag model. Both quantities are normalized to the pure
Harrison-Zel'dovich radiation amplitude. On the horizontal axis the wave number
$k$ is represented by the CDM mass contained in a sphere of radius $\pi/k$.} 
\end{figure}

For the bag model the transfer functions are shown in Fig.~\ref{fig5}. Both
transfer functions show huge peaks on small scales. The different scales
$k=2\pi/\lambda$ are represented by the invariant CDM mass contained in a
sphere with radius $\lambda/2$,  
\begin{equation}
M^{\rm CDM}(\lambda^{\rm phys})\equiv \frac{4 \pi}{3}\rho^{\rm CDM}
\left(\frac{\lambda^{\rm phys}}{2}\right)^3 \ ,
\end{equation}
assuming  that $\Omega_{\rm CDM} \approx 1$ today. The largest scales 
in Fig.~\ref{fig5} correspond to the
horizon at $T_{\rm f}=T_{\star}/10$. The CDM curve also shows the
logarithmic growth of subhorizon scales of CDM in a radiation dominated
universe. The CDM peaks lie on top of this logarithmic curve.

The peak structure starts at a scale $\approx 10^{-8} M_{\odot}$ in CDM
mass. This scale corresponds to the horizon scale at the QCD transition,
$M_{\rm H} \equiv {4 \pi \over 3} \rho^{\rm CDM} (R_{\rm H}/2)^3$. The
radiation energy inside the horizon at $T_{\star}$ is $\sim 1 M_{\odot}$, but
it is redshifted as $M^{\rm RAD}(a) \sim (a_{\rm equality}/a) M^{\rm
CDM}$. Scales which are above the horizon at the QCD transition are not
affected in accordance with the general proof in the previous chapter,
Eq.~(\ref{super}). For scales below  $M_1^{\rm CDM} \approx 9 \times 10^{-9}
M_{\odot}$ the radiation peaks grow linearly in wave number.  
This linear growth comes from the fact that the vanishing sound speed during 
the QCD transition implies a vanishing restoring force in the acoustic 
oscillations on subhorizon scales. Therefore, the radiation fluid falls freely
during the transition, with a constant velocity given at the beginning of the
transition. The density contrast $\delta^{\rm RAD}$ grows linearly in time
with a slope $k$. CDM is moving in an external potential provided by the
dominant radiation fluid, and is pushed by the strong increase in the
gravitational force during the transition. The highest peaks have 
$k/k_1 \sim 10^4$, because on smaller scales the acoustic oscillations are
damped away by neutrino diffusion already before the QCD transition 
(see Sec. IV).

\begin{figure}
\begin{center}
\epsfig{figure=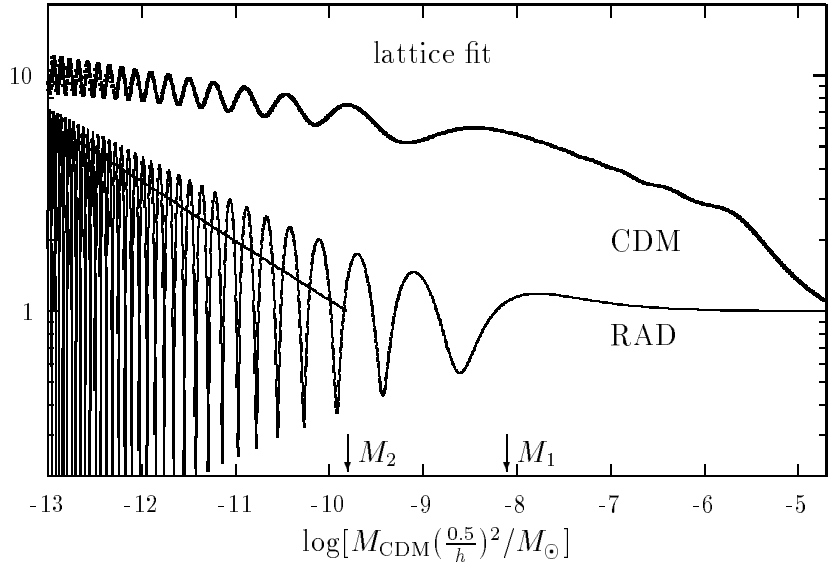,width=0.67\linewidth}
\end{center}
\caption{\label{fig6}
The same as Fig.~\ref{fig5}, but for our lattice QCD fit. The
straight line denotes the asymptotic envelope for small scales 
$\propto k^{3/4}$.} 
\end{figure}

For our fit to lattice QCD the transfer functions ${\mathcal T}^{\rm RAD},
{\mathcal T}^{\rm CDM}$ are shown in Fig.~\ref{fig6}. The peak structure
starts at the horizon scale $M_1$, but the asymptotic growth on small
scales is different. It grows $\propto k^{3/4}$ starting at a scale
$k_2$ which corresponds to an invariant CDM mass of $M_2 = 2 \times
10^{-10} M_{\odot}$. The asymptotic envelope for small scales is indicated by
the straight line in Fig.~\ref{fig6}. The highest peaks show an amplification
of a factor $400$ and correspond to wave numbers $k/k_2\sim 3 \times 10^3$.
The smaller amplification compared to the bag model is due to the decrease
of the radiation fluid velocity before the transition. This happens 
because the sound speed drops below $1/\sqrt{3}$ already before the 
transition. The values for
$M_2$ and the exponent of asymptotic growth depend on our specific choice of 
the latent heat ($R_L=0.2$) and the thermodynamic behavior at $T >
T_{\star}$, parameterized by $\gamma$ ($\gamma=1/3$). In the next chapter we
will show, that the exponent of the asymptotic envelope for $\gamma \in (0,1)$
varies between $1/2$ and $1$. 
 
The processed spectrum for a crossover, Fig.~\ref{fig7}, shows a 
similar behavior as for the bag model and the lattice fit on superhorizon and 
horizon scales. The peak structure starts at $M_1$, but on subhorizon scales
there are no peaks. The level of the subhorizon transfer function for the
radiation fluid is reduced to $0.83$. This comes from the damping of 
the acoustic oscillations during the time with $c_s^2 \ne 1/3$. 

\subsection{Analytic solution for the radiation fluid}

The time evolution for subhorizon modes, $\lambda^{\rm phys} \ll R_{\rm
H}=H^{-1}$, at the transition can be solved analytically. For the dynamics of
the radiation fluid (QCD, photons, leptons) one can neglect cold dark matter
(CDM), since $\rho^{\rm CDM}/\rho^{\rm RAD} = a/a_{\rm equality} \approx
10^{-8}$. The transition time is short compared to the Hubble time at the
transition, $(t_+ - t_-) < t_{\rm H}=H^{-1}$. For subhorizon modes we can
neglect gravity during the whole transition as we will show below. The damping
terms in the continuity equation (\ref{C2}) and Euler equation (\ref{E2}) are
absent in the purely radiation dominated regime. 
During the transition the damping terms can be neglected in view of 
the huge amplification for a first-order phase transition. Eqs.~(\ref{C2})
and (\ref{E2}) now read:  
\begin{eqnarray}
\label{psihat}
  \delta' -k \hat{\psi}  = 0 \\\nonumber
  \hat{\psi}' + c_s^2 k\delta=0 \ .
\end{eqnarray}
Written as a second order differential equation for $\delta$, this is just an
oscillator equation 
\begin{equation}
\label{oscillator}
  \delta '' + \omega^2 \delta =0 \ ,
\end{equation}
with the time-dependent frequency $\omega(\eta) \equiv k c_s(\eta)$.
 
\subsubsection{Bag model}

We first discuss the origin of large amplifications for small scales in the
bag model. Before and after the QCD transition the radiation fluid
makes standing acoustic oscillations in each mode $k$ with sound speed $c_s^2 =
1/3$ and with amplitude $A_{\rm in}$ and $A_{\rm out}$ for the incoming and
outgoing solution respectively, see Fig.~\ref{fig8}. The incoming solution for
the density contrast $\delta$ and the peculiar velocity $\sqrt{3}\hat{\psi}$
reads  
\begin{eqnarray}
 \delta &=&-A_{\rm in} \cos{(\omega(\eta-\eta_-)+\varphi_-)} \\ \nonumber
 \sqrt{3}\hat{\psi}&=&A_{\rm in} \sin{(\omega(\eta-\eta_-)+\varphi_-)} \ .
\label{bagin}
\end{eqnarray}
This solution is valid until the beginning of the transition at
$\eta=\eta_{-}$, and $\varphi_-$ denotes the phase of the oscillation at
$\eta_-$. During the transition the sound speed is zero. There are no
restoring forces from pressure gradients and the radiation fluid falls
freely. Since the duration of the transition is short compared to the Hubble
time  $\Delta t\equiv (t_+ - t_-) < H^{-1}$, gravity is negligible during this
free fall. If we estimate the contribution of the gravity term in the Euler
equation (\ref{E2}) during the transition we get $(t_+ - t_-)
\partial_t\hat{\psi}_- = 
{\cal O}(\Delta t k_{\rm phys}) \alpha_- = 
{\cal O}[(H\Delta t)(H/k_{\rm phys})]\delta_- = 
{\cal O}[(H\Delta t)(H/k_{\rm phys})] \hat{\psi}_- \ll
\hat{\psi}_-$ except for an initial phase 
$\varphi_- = 0$ mod $\pi$ (which leads to a dip in the spectrum).  
Thus, the fluid is moving
inertially in the sense of Newton, the velocity stays constant, and the density
contrast grows linearly in time 
\begin{eqnarray}
 \delta &=& \delta_-+k(\eta-\eta_-) \hat{\psi}_- \ , 
 \label{freefall} \\ \nonumber
 \sqrt{3}\hat{\psi}&=&\sqrt{3}\hat{\psi}_- \ , 
\end{eqnarray}
where $\sqrt{3}\hat{\psi}_-=A_{\rm in} \sin{(\varphi_-)}$ is the peculiar
velocity at $\eta_-$. 

Since we have no jumps in pressure $p_0$, the density
contrast $\delta$ and the fluid velocity $\hat{\psi}$ stay continuous
throughout the whole transition, in particular at the matching points of the
different regimes. Gravity remains negligible during the entire transition
since the growth of $\hat{\psi}$ due to gravity at the end of the transition
can be estimated by $\hat{\psi}_+ - \hat{\psi}_- = 
{\cal O}(\Delta t k_{\rm phys} \alpha_+) = 
{\cal O}[H\Delta t(H/k_{\rm phys}) \delta_+]=
{\cal O}[(H \Delta t)^2 \hat{\psi}_-] \ll \hat{\psi}_-$, in the limit under
consideration. At the end of the transition this solution has to be
joined to the pure radiation dominated regime for $T\le T_{\star}$. Since the
amplitude of the density contrast grows linearly during the transition, the
final amplitude $A_{\rm out} = A_+$ is enhanced linearly in $k$,
modulated by the incoming phase  
\begin{equation}
\label{k1}
  \left(\frac{A_{\rm out}}{ A_{\rm in}}\right)^2 \!\!\!= 
\left({k \over k_1}\right)^2 \sin^2{(\varphi_-)} \ ,
\end{equation}
with $k_1\equiv \sqrt{3}/\Delta \eta$, $\Delta \eta \equiv \eta_+-\eta_-$.
The envelope of the linearly growing peak structure for subhorizon scales
starts at the scale $k_1$ which corresponds to a CDM mass of $M_1 = 9 \times
10^{-9} M_{\odot}$. 

\subsubsection{Lattice QCD}

For our fit to lattice QCD, Eq.~(\ref{fit}), the time evolution separates into
four regimes. First we have a WKB regime,
where the sound speed is slowly changing compared to the frequency of the
oscillation, i.e. the WKB condition $|dc_s/d\eta|/c_s \ll \omega = k
c_s$ is  satisfied. The WKB solution is 
\begin{eqnarray}
\delta &=& - \frac{A_{\rm in}}{(3 c_s^2)^{1/4}} \cos{\int_0^\eta \omega(\eta')
d\eta'}\\ 
\nonumber 
\sqrt{3}\hat{\psi} &=& A_{\rm in}(3 c_s^2)^{1/4} \sin{\int_0^\eta \omega(\eta')
d\eta'} \ ,
\end{eqnarray}
where $A_{in}$, as before, denotes the incoming subhorizon amplitude in a radiation
dominated universe. The decreasing sound speed in the WKB regime leads to an
increase in the density contrast and a decrease in peculiar velocity compared
to the purely radiation dominated case. Close to $\eta_{-}$ the WKB condition
can no longer be satisfied: Since $c_s$ decreases by some power in
$(\eta_{-}-\eta)$, Eq.~(\ref{cseta-}), the left hand side of the WKB condition
diverges like $1/(\eta_{-}-\eta)$, whereas the right hand side for any fixed
$k$ goes to zero. Therefore, the WKB solution for a given mode is valid up to a
certain time $\tilde{\eta}(k) < \eta_{-}$. $\tilde{\eta}(k)$ approaches
$\eta_{-}$ for increasing $k$. 

Close to $\eta_-$ there is a second regime
(overlapping with the WKB regime for large enough k), where $c_s$ can be
approximated by the power law of Eq.~(\ref{cseta-}),
\begin{eqnarray}
\label{powerlaw}
c_s &=& C \left( {\eta_- -\eta \over \eta_-}\right)^{{1\over 2\beta}-1} \\
\nonumber
\beta &\equiv& \frac{\gamma}{\gamma+1} \ ,
\end{eqnarray}
where $\beta$ is defined for convenience below. Inserting this power law
into the oscillator equation (\ref{oscillator}) gives a Bessel differential
equation. The new variable $z$ is the argument of the cosine in the WKB
expansion, 
\begin{equation}
z \equiv k\int_{\eta}^{\eta_-} c_s(\eta')d\eta'= 2 C \beta (k
\eta_-)\left( {\eta_- -\eta \over \eta_-}\right)^{\frac{1}{2\beta}} \ .
\end{equation}
The index $\bar{\beta}$ of the Bessel functions is obvious from the ratio of
the two fundamental solutions $J_{\bar{\beta}}(z)/N_{\bar{\beta}}(z) \propto
z^{2 \bar{\beta}} \propto (\eta_- - \eta)^{\bar{\beta}/\beta}$. Since the two
fundamental solutions for $c_s \rightarrow 0$ are $\delta \rightarrow$
const and $\delta \propto (\eta_- - \eta)$, it follows that $\bar{\beta} =
\beta$. The prefactor must be $z^\beta$ to make the second fundamental solution
constant,
\begin{equation}
\label{regime2}
  \delta = z^\beta \left[-B_1 J_{\beta}(z) + B_2 N_{\beta}(z)\right] \ .
\end{equation}
The peculiar velocity follows via $\hat{\psi} = \delta'/k $, see
Eq.~(\ref{psihat}). Matching the WKB solution with the
Bessel solution in the overlap region $z\gg1$, where  $J_{\beta}(z), 
N_{\beta}(z) \rightarrow \sqrt{2/(\pi z)} \cos[z-(\pi/2)(\beta +1/2,3/2)]$,
gives the normalization $B^2 \equiv
B_1^2 + B_2^2$,
\begin{equation}
  B=A_{\rm in}{\sqrt{\pi\beta}\over3^{1/4}}(2
  C\beta)^{-\beta}(k \eta_-)^{1/2-\beta} \ .
\end{equation} 

At the beginning of the first-order transition, $\eta = \eta_-$, the
Bessel solution (second regime) is matched to the third regime, i.e. the free
fall regime with $ c_s^2 = 0 $. As in the bag model, peaks in the transfer
function for large k are obtained for the mixing angles $\varphi$ in the
Bessel solution ($B_1=B \cos{\varphi}, B_2=B \sin{\varphi}$) which give the
maximal $|\hat{\psi}|$ at the matching point. With $J_{\beta}(z)\rightarrow
(z/2)^{\beta}/\Gamma(\beta+1)$ for $z \ll 1$ and $N_{\beta}(z)= [J_{\beta}(z)
\cos(\beta \pi) - J_{-\beta}(z)]/\sin(\beta \pi)$ we find
\begin{equation}
\label{vlattice}
  \hat{\psi}^{\rm max}_-=A_{\rm in} \frac{\sqrt{\pi\beta}}{3^{1/4}}
           \frac{(C\beta)^\beta}{\Gamma(\beta+1)\sin{\beta\pi}}
           \left(k \eta_-\right)^{\beta -\frac{1}{2}} \ .
\end{equation}
The evolution in the third regime, $c_s^2=0$, is as before. The restoring
force in the oscillator equation, Eq.~(\ref{oscillator}), vanishes, the
fluid moves with constant peculiar velocity, $\hat{\psi}(\eta)=
\hat{\psi}_{-}$, and the density contrast grows linearly in time. 

After the transition ($\eta>\eta_{+}$, fourth regime) the universe is
radiation dominated ($c_s^2=1/3$) and the solutions are oscillations with
amplitude $A_{\rm out} = A_{+}$. The amplification of peaks in the
outgoing amplitude $A_{\rm out}$ is given by  $\left.A_{\rm out}\right|_{\rm
peaks} = \hat{\psi}^{\rm max}_{-} k \Delta \eta$,   
\begin{equation}
  \left. \frac{A_{\rm out}}{A_{\rm in}} \right|_{\rm peaks} \!\!\! =
  \frac{\sqrt{\pi\beta}}{3^{1/4}} \frac{(C\beta)^\beta}{\Gamma(\beta+1)
  \sin{\beta\pi}} \left(k \eta_-\right)^{\beta -\frac{1}{2}}   
   k \Delta \eta. 
\end{equation}
The dependence on the latent heat is encoded in the duration $\Delta \eta$ of
the transition. $C$ and $\beta$ are determined by the normalization and the
power law for $c_s(\eta)$ shortly before the onset of the first-order phase
transition, Eq.~(\ref{powerlaw}). Replacing $\beta$ by $\gamma$ (= the
exponent in the QCD fit to the entropy density) the amplification factor is 
\begin{equation}
  \left. \frac{A_{\rm out}}{A_{\rm in}} \right|_{\rm peaks} \!\!\! =
  \left(k\over k_2\right)^{\frac{3\gamma+1}{2(\gamma+1)}} 
\end{equation} 
for large $k$. Since $0<\gamma<1$ the exponent of the power law varies between
$1/2$ and $1$. The normalization of the peak amplification for large $k$ is
encoded in $k_2$. In our fit to lattice QCD we take $\gamma = 1/3$, hence
$\beta =1/4$
\begin{equation}
  \hspace{-1pt} \left. A_{\rm out}\over A_{\rm in}\right|_{\rm peaks} \!\!\! = 
  \left(k\over k_2\right)^{\frac34}. 
\end{equation}
For $R_L = 0.2$, $k_2$ corresponds to a CDM mass of $M_2 = 2 \times 10^{-10}
M_{\odot}$. The spectrum for this case together with the asymptotic
envelope is plotted in Fig.~\ref{fig6}.

\subsubsection{Crossover}

In the case of a crossover, the amplification is occurring for scales
around the Hubble radius at the transition only. Subhorizon scales always
stay in the WKB-regime and therefore the spectrum is flat for these scales. 
However, the amplitude for subhorizon scales is damped during the 
phase transition. The same damping occurs in the case of a first-order
phase transition. It has been neglected in the analytic discussion, since 
it is a small correction.
The time evolution of the density contrast of subhorizon scales, including 
damping terms, reads  
\begin{equation}
   \delta'' + (1+3c_s^2-6w)\frac{a'}{a}\delta'+\omega^2(\eta) \delta=0 \ .
   \label{damposc} 
\end{equation}
The friction term in this equation can be eliminated if we write the density
contrast as $\delta =: \frac{\sqrt{1+w}}{a'} u(\eta)$. Inserting this 
ansatz
into Eq.~(\ref{damposc}), using the Friedmann equation and $w'=3 a H
(1+w)(w-c_s^2)$, we obtain an undamped oscillator equation with frequency
$\omega(\eta) = k c_s(\eta)$ for the quantity $u(\eta)$. For a crossover the
sound velocity never vanishes, and subhorizon modes always satisfy the
WKB-condition. Therefore, we obtain for the density contrast and the peculiar
velocity 
\begin{eqnarray}
\delta &=& - \frac{A_{\rm in}}{(3 c_s^2)^{1/4}} \frac{a'_{\rm in}}{a'} 
            \frac{\sqrt{3(1+w)}}{2} \cos{\int \omega(\eta) d\eta}\\
\nonumber 
\hat{\psi} &=& A_{\rm in}(3 c_s^2)^{1/4} \frac{a'_{\rm in}}{a'} 
             \frac{\sqrt{ 3(1+w)}}{2} \sin{\int \omega(\eta) d \eta}.
\end{eqnarray}
The reduction of the amplitude of subhorizon scales is given by 
\begin{equation}
   \frac{ A_{\rm out}}{ A_{\rm in}} 
   =\frac{(H a^2)_{\rm in}}{(H a^2)_{\rm out}}=
    \left(\frac{g_+}{g_-}\right)^{1/6}. 
\end{equation}
Therefore the subhorizon amplitudes are reduced to 83 \% of their initial
value. This damping is a general feature appearing whenever the number of
degrees of freedom is reduced. Also in the case of a first-order transition,
this damping factor should be taken into account. However, it changes the huge
amplifications merely by the constant factor 0.83.

\subsection{Analytic solution for CDM}

Cold dark matter (CDM) is assumed to be the major matter content of the
universe today.
At the time of the QCD transition, however, the contribution of CDM 
to the total energy density was negligible, $\rho^{\rm CDM}/\rho^{\rm RAD} =
a/a_{\rm equality} \approx 10^{-8}$. We consider a type of CDM which is
non-relativistic ($p \ll \rho$) at the QCD transition and which is only
coupled via gravity to the radiation fluid. CDM moves in the external
gravitational potential provided by the dominant radiation fluid. During the
QCD transition, the big amplifications of the density contrast in the radiation
fluid $\delta^{\rm RAD}$ leads to a big amplification in the gravitational
potential, i.e., the force term in the Euler equation for CDM. The CDM is 
accelerated to higher velocities at the end of the transition. Therefore, we
also get peaks and dips in the cold dark matter fluid.

\begin{figure}
\begin{center}
\epsfig{figure=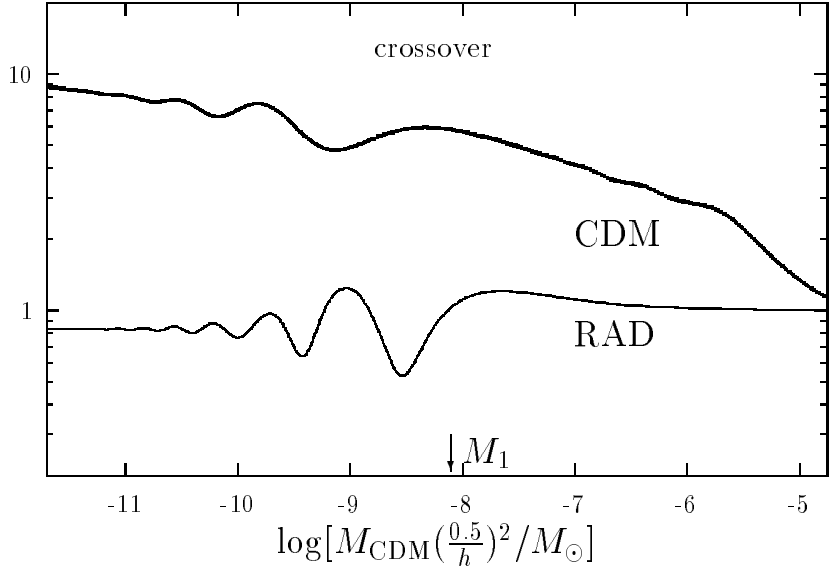,width=0.67\linewidth}
\end{center}
\caption{\label{fig7}
The same as Fig.~\ref{fig5}, but for a QCD crossover.} 
\end{figure}

\begin{figure}
\begin{center}
\epsfig{figure=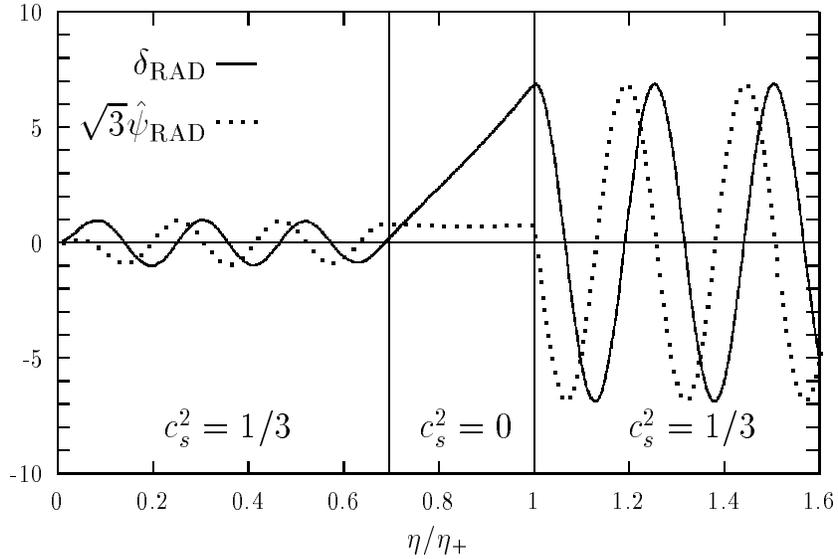,width=0.67\linewidth}
\end{center}
\vspace{-7pt}
\caption{\label{fig8}
The time evolution of the density contrast, $\delta^{\rm RAD}$,  
and the peculiar velocity, $\frac34 \hat{\psi}^{\rm RAD}$,
of the radiation fluid for the mode $k/k_1=7$ in uniform expansion gauge. 
During the QCD transition in the bag model --- 
marked by the 2 vertical lines --- the velocity stays
approximately constant and the density contrast grows linearly.
The amplitude is normalized to $1$ long before the transition.} 
\end{figure}

\begin{figure}
\begin{center}
\epsfig{figure=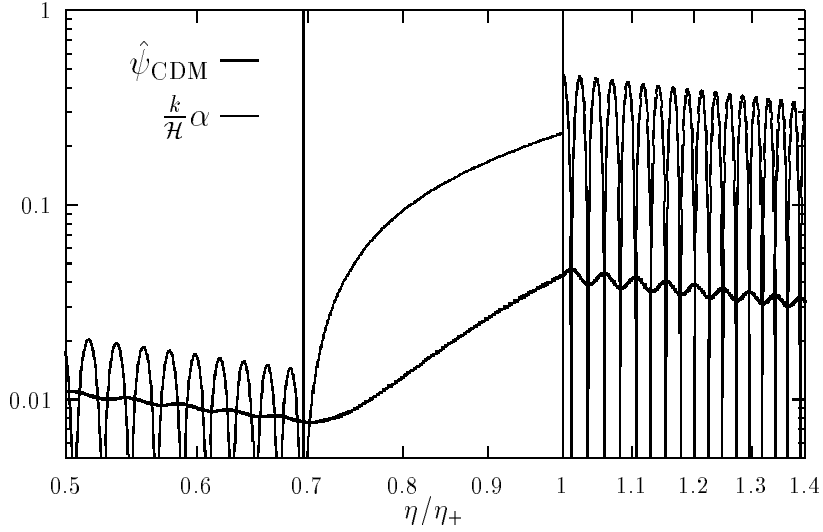,width=0.67\linewidth}
\end{center}
\vspace{-7pt}
\caption{\label{fig9}
The time evolution of the peculiar velocity, $\hat{\psi}^{\rm CDM}$,  
of the kinetically decoupled cold dark matter fluid (axions or primordial 
black holes) and the external gravitational potential gradient $\frac{k_{\rm
phys}}{H}\alpha$ for the mode $k/k_1=37$ in uniform expansion gauge. $\alpha$
is provided by the radiation fluid. During the QCD transition in the bag model
--- marked by the 2 vertical lines --- the cold dark matter fluid is
accelerated by the gravitational force.}  
\end{figure}

\begin{figure}
\begin{center}
\epsfig{figure=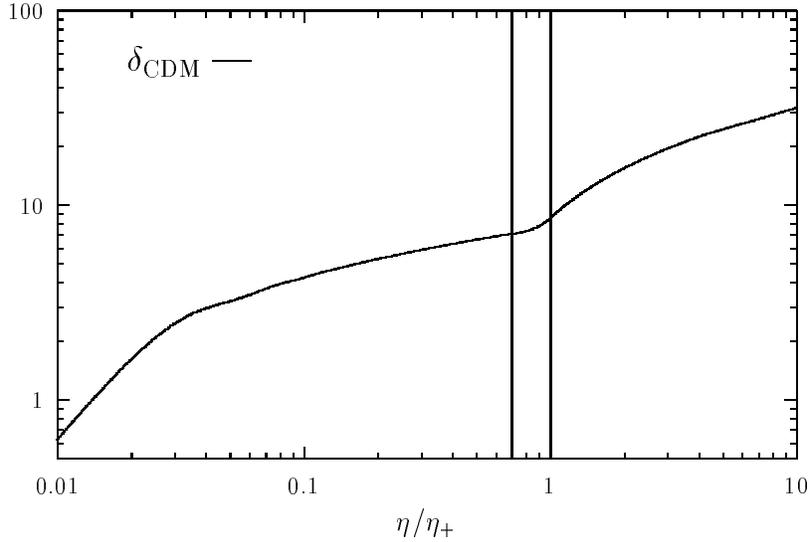,width=0.67\linewidth}
\end{center}
\vspace{-7pt}
\caption{\label{fig10}
The time evolution of the CDM density contrast, $\delta^{\rm CDM}$, 
corresponding to the CDM velocity of Fig.~\ref{fig9}. 
The major amplification of $\delta^{\rm CDM}$ is due to the higher 
peculiar velocity at the end of the transition.} 
\end{figure}

The subhorizon evolution of CDM in a purely radiation dominated universe is
just inertial motion. This can be seen from the Euler equation (\ref{E2}):
The sound velocity of cold dark matter is zero, $ c_s^2\equiv 0$, and the
gravitational potential $\alpha$ is given using the Poisson equation (\ref{R2})
for the dominant density contrast $\delta^{\rm RAD}$ Eq.~(\ref{bagin}). 
In the leading order in $x=k/{\mathcal H}$, the
gravitational force can be neglected and the subhorizon evolution of the CDM
velocity is obtained:
\begin{equation}
\hat{\psi}^{\rm CDM} = C \frac{1}{x}.
\label{CDMinitial}
\end{equation}
The velocity of CDM in a radiation dominated universe just redshifts to
zero on subhorizon scales, see Fig.~\ref{fig9}. $C$ is an integration constant
of order $A_{\rm in}$. The corresponding evolution
of the density contrast $\delta^{\rm CDM}$ follows from the continuity
equation (\ref{C2}), 
\begin{equation}
\label{subCDM}
\delta^{\rm CDM} = C \ln{x} + D.
\end{equation}
This logarithmic growth of $\delta^{\rm CDM}$ can be seen in
Fig.~\ref{fig10} before and after the transition. The shape of $\delta^{\rm
CDM}$ can also be seen in the transfer functions
Figs.~\ref{fig5},\ref{fig6},\ref{fig7} on scales above the horizon scale $M_1$.

During a first-order QCD transition the evolution of the gravitational
potential is dominated by the linear growth in time of $\delta^{\rm RAD}$,
Eq.~(\ref{freefall}). From the Poisson equation (\ref{R2}) follows that the gravitational potential grows linearly in time 
\begin{equation}
\alpha = - \frac32 {\mathcal H}(\eta-\eta_-) \left(\frac{\mathcal H}{k}\right)
\hat{\psi}_-^{\rm RAD},
\end{equation}
see Fig.~\ref{fig9}. As before, we will take the transition time to be short
compared to the Hubble time $\Delta t <H^{-1}$, hence
$\Delta \eta <{\mathcal H}^{-1}$. The gravitational force in the Euler
equation  (\ref{E2}) and hence the fluid acceleration for CDM also grows
linearly in time during the transition. The fluid velocity grows quadratically
\begin{equation}
\label{psiCDM}
\hat{\psi}_+^{\rm CDM} = \frac34 \left[{\mathcal H}\Delta\eta\right]^2
\hat{\psi}_-^{\rm RAD} +\hat{\psi}^{\rm CDM}_-.
\end{equation}
The initial CDM velocity $\hat{\psi}^{\rm CDM}_-$ is given by
Eq.~(\ref{CDMinitial}). On scales well below the horizon the velocity 
is redshifted to small values and can be neglected. However, on subhorizon 
scales closer to the horizon the initial velocity dominates, because 
$[{\mathcal H}(\Delta\eta)]^2 \approx 0.1 (0.01)$ for the bag model (lattice
fit) is small. According to the continuity equation (\ref{C2}) the density
contrast $\delta^{\rm CDM}$ grows cubically
\begin{equation}
\delta^{\rm CDM}_+ = 
\delta_-^{\rm CDM} +
\frac14 \left(\frac{k}{\mathcal H}\right)\left[{\mathcal
H}\Delta\eta\right]^3 \hat{\psi}_-^{\rm RAD}.
\end{equation}
Due to the cube the growth of $\delta^{\rm CDM}$ during
the QCD transition is small. 

The major amplification effect comes from the 
higher velocity at the end of the transition (see Fig.~\ref{fig9}), 
which leads to 
\begin{equation}
\label{growthCDM}
\delta^{\rm CDM}(\eta) =  \hat{\psi}^{\rm CDM}_+ k\eta_+ 
\ln({\eta\over\eta_+}) + \delta^{\rm CDM}_+. 
\end{equation}
The amplification of the density contrast during the transition is negligible
compared to the enhancement of the velocity. This velocity enhancement during
the transition leads to an additional logarithmic growth of $\delta^{\rm CDM}$
after the transition, 
\begin{equation}
\delta^{\rm CDM}(\eta) = 
\left[\frac34\left({\Delta\eta\over \eta_+}\right)^2 
\hat{\psi}_-^{\rm RAD}\right] 
k\eta_+ \ln({\eta\over\eta_+}) + \delta^{\rm CDM}_+.
\end{equation}
Here we have used (\ref{psiCDM}) to obtain the leading subhorizon 
approximation. This additional logarithmic growth of $\delta^{\rm CDM}$ is most clearly seen in Fig.~\ref{fig10}. 

The number of CDM peaks in the numerical result, Figs.~\ref{fig5}-\ref{fig7},
is only half of the radiation peaks. This comes from the fact that the peaks in
$\delta^{\rm CDM}$ lie on top of the logarithmic curve, which is still the
main contribution  at the scales plotted. 
The dips correspond to modes, with a
maximal negative amplification.  

The jump in the gravitational potential at the end of the QCD transition in
Fig.~\ref{fig9} follows from the Poisson equation (\ref{R2}). Since $\delta$ is
continuous, but $c_s^2$ jumps, the right hand side abruptly changes by a
factor 2. Therefore $\alpha$ jumps by the same factor.

Let us compare the `QCD peaks' in the CDM spectrum with the CDM spectrum 
without phase transition. In the limit $\rho^{\rm CDM}/\rho^{\rm RAD}
\ll 1$ the exact solution to Eqs.~(\ref{C1}) -- (\ref{R1}) in the radiation
dominated universe without transition reads ($x = k/{\cal H}$):
\begin{eqnarray}
\delta^{\rm CDM} &=& {3 A_{\rm in}\over 2}\left[ \ln({x \over \sqrt{3}}) 
+ \gamma_{\rm E} - \frac12  - {\rm Ci}({x\over \sqrt{3}})+ \right. \nonumber \\
& & \left. \hspace{1.2cm} + {3\over x^2 + 6}
\left({x\over\sqrt{3}} \sin({x\over\sqrt{3}}) + \cos({x\over\sqrt{3}})\right)
\right] \\
\hat{\psi}^{\rm CDM} &=& {3 A_{\rm in}\over 2 x}\left[ 1 - 
{6\over x^2 + 6}
\left({x\over\sqrt{3}} \sin({x\over\sqrt{3}}) + \cos({x\over\sqrt{3}})\right)
\right] \ .
\end{eqnarray}
In the subhorizon limit this solution reduces to 
\begin{eqnarray}
\delta^{\rm CDM} &=& {3 A_{\rm in}\over 2}
\left[\ln({x\over\sqrt{3}}) + 
\underbrace{\gamma_{\rm E} - \frac12}_{\approx 0.077} \right] \\
\hat{\psi}^{\rm CDM} &=& {3 A_{\rm in}\over 2} \frac 1x \ ,
\end{eqnarray}
which fixes the constants $C$ and $D$ in Eqs.~(\ref
{CDMinitial}) and (\ref{subCDM}).

We are now able to calculate the enhancement factor
\begin{equation}
\label{enhancement}
{\cal E}(\eta) \equiv \left[\delta_{\rm transition}\over 
\delta_{\rm no\ transition}\right]^{\rm CDM}(\eta) \ .
\end{equation}
Let us evaluate ${\cal E}$ at equality, i.e., $\eta_{\rm eq}/\eta_+ \approx 
2\times 10^8$: 
\begin{equation}
{\cal E}(\eta_{\rm eq}) \approx 1 + \left[\frac12
\left({\Delta\eta\over \eta_+}\right)^2
\hat{\psi}_-^{\rm RAD}\right] k\eta_+ {\ln({\eta_{\rm eq}\over\eta_+})\left/
\ln({k \eta_{\rm eq}\over \sqrt{3}}) \right. } \ .
\end{equation}
The enhancement at the smallest scales $M^{\rm CDM} \sim 10^{-20} M_\odot$ 
to which our calculation applies and at some intermediate scale $M^{\rm CDM}
\sim 10^{-15} M_\odot$ reads
\begin{equation}
{\cal E}(\eta_{\rm eq}) \approx \left\{ 
\begin{array}{ll}
600 (20) & \mbox{bag\ model} \\
10  (2)  & \mbox{lattice\ fit}
\end{array} \right\} \quad 
\mbox{for\ } 10^{-20} (10^{-15}) M_\odot \ ,
\end{equation}
where we have used Eq.~(\ref{vlattice}) for
the case of the lattice fit.
This shows that both models lead to large enhancements of the CDM density
contrast at equality for small enough scales. For the lattice fit the
enhancement becomes important at scales below $M^{\rm CDM} \sim 10^{-15}
M_\odot$. 
  
\section{Implications --- clumps in CDM}

\subsection{Fate of the peaks in the radiation fluid}

Before discussing CDM, let us show that the large peaks in the density
spectrum of the hadron-lepton-photon fluid do not lead to any observable 
consequences.
 
\subsubsection{Collisional Damping at Neutrino Decoupling}

The acoustic oscillations in the radiation fluid get damped by neutrino
diffusion at the time of neutrino decoupling. This damping is analogous to
Silk damping at photon decoupling. The muon resp. tau neutrinos, which are
coupled to the RAD fluid via neutral current interactions only, decouple at
$T^{\rm dec}_{\nu_\mu \; \nu_\tau} \sim 2.2$ MeV from the Hubble scale $R_{\rm
H}$, which follows from Ref.~\cite{Heckler}. The electron neutrinos interact
by charged and neutral currents and decouple slightly later, $T^{\rm
dec}_{\nu_e} \sim 1.4$ MeV. By the time of neutrino decoupling at the Hubble
scale all inhomogeneities in the RAD fluid on scales below $\approx 10^{-6}
M_{\odot}$ in CDM mass are wiped out by neutrino diffusion (cf.~the QCD
horizon scale is $10^{-8} M_\odot$ in CDM mass), as shown below,
Eq.~(\ref{nudamping}). Therefore our QCD peaks cannot affect big bang
nucleosynthesis (BBN).         

It is important to distinguish the total decoupling of neutrinos,
i.e. neutrino decoupling at the Hubble scale, when the neutrino
scattering rate $\Gamma$ is less than the Hubble rate, $\Gamma_{\nu}/H<1$,
from the neutrino decoupling with respect to a 
certain mode given by  $\Gamma_{\nu}/\omega_{\rm phys}<1$, when the typical
neutrino scatters less than once during an acoustic oscillation time of one
particular mode. The mode-dependent decoupling temperature $T_{\nu}^{\rm
dec}(k)$ is related to the total decoupling temperature by
\begin{equation}
{T^{\rm dec}_{\nu_i}(k) \over T^{\rm dec}_{\nu_i}(H)} \approx \left({ c_s
k_{\rm phys}\over H}\right)^{1/4}_{T=T^{\rm dec}_{\nu_i}(H)} \ .
\end{equation}
This follows because the neutrino interaction rates $\Gamma_{\nu}$ are
proportional to $T^5$ and $k_{\rm phys} \propto 1/a \propto T$, hence
$\Gamma_{\nu}/\omega_{\rm phys} \propto T^4$.

To compute the damping of acoustic oscillations in the radiation fluid by
neutrino diffusion we follow Weinberg \cite{Weinberg}. For a radiation fluid
shear viscosity is dominant,  bulk viscosity vanishes  and heat conduction is negligible. The shear viscosity is given by 
\begin{equation}
 \eta_{\rm visc}={4 \over 15} \sum_i \rho_{\nu_i} \tau_{\nu_i} \ .
\end{equation}
$\rho_{\nu_i}$ denotes the energy density of a neutrino species,
$\tau_{\nu_i}$ is the typical collision time. In the subhorizon limit the
Navier-Stokes equation and the continuity equation give 
\begin{equation}
  \delta'' + {k_{\rm phys} \over \rho_{\rm tot}} \eta_{\rm visc} k \delta' +
  \omega^2 \delta = 0, 
\end{equation}
a damped oscillator. The damping factor for the mode $k$ at a given conformal
time $\eta$ is
\begin{equation}
\label{dampingfactor}
D(k,\eta)= \exp{[-{1 \over 2}\int_0^{\eta_{\rm max}}
 (k_{\rm phys} / \rho_{\rm tot}) \eta_{\rm visc} k d \eta']},
\end{equation}
Here the upper limit of the integral is the conformal time $\eta_{\rm max} =
{\rm Min} \left[\eta,\eta_{\rm dec}(k)\right]$, because 
collisional damping of the $\gamma$-$\ell^{\pm}$-hadron fluid by neutrinos
ceases at decoupling of the mode $k$ considered. The damping per oscillation
is largest for $\omega \tau_{\nu} \equiv \omega/\Gamma_{\nu} \approx 1$,
i.e. immediately before neutrino decoupling for a given  mode. But note
that subhorizon modes get strongly damped long before the mode decouples from
neutrinos, because a weak damping per oscillation is compensated by many
oscillations per Hubble time.

 For a first application we ask what modes are already damped by the time of
the QCD transition. At the QCD transition, $T=T_{\star}$, the interaction rates
for electron and muon neutrinos (resp. antineutrinos) with the leptons are the
same, $\Gamma_{\nu_e}=\Gamma_{\bar{\nu}_e}=\Gamma_{\nu_\mu}=
\Gamma_{\bar{\nu}_\mu} =3.1 G_{\rm F}^2 T^5$ \cite{Heckler},  since electrons
and muons are still  relativistic. The $\tau-$neutrinos interact only via
neutral currents with the leptons and have a lower interaction rate,
$\Gamma_{\nu_\tau}= \Gamma_{\bar{\nu}_\tau} = 0.6 G_{\rm F}^2 T^5$.  
Evaluating the damping integral, Eq.~(\ref{dampingfactor}), at $T=T_\star$ we
find that the damping factor $D(k,T_\star)$ is $<1/e$ for $(k_{\rm
phys}/H)_{T_{\star}} > 10^4$, which corresponds to $M_{\rm CDM} <10^{-20}
M_{\odot}$, i.e. acoustic oscillations on these small scales are wiped out
before the QCD transition. Therefore no peaks in the RAD or in the CDM
transfer function can develop below this scale. This small-scale cut-off is
independent of the bubble separation scale.

Next we consider $T<T_{\nu_e}^{\rm dec}(H) \approx 1.4$ MeV and evaluate the
final damping factor $D(k)$.
At the time of neutrino decoupling we take a purely radiation dominated
universe, consisting of $\gamma$, $e^{\pm}$ and $\nu's$ only to evaluate the
damping. Muons have disappeared since $m_{\mu} \gg T$ and since $\tau_{\mu}
\ll t_{\rm H}$. The interaction rate for electrons is still given by charged
and neutral currents, $\Gamma_{\nu_e}=\Gamma_{\bar{\nu}_e}=1.3 G_{\rm F}^2
T^5$, whereas muon and tau neutrinos have the same lower (neutral current)
rate, $\Gamma_{\nu_\mu}=\Gamma_{\bar{\nu}_\mu}=\Gamma_{\nu_\tau}=
\Gamma_{\bar{\nu}_\tau}=0.3 G_{\rm F}^2 T^5$ \cite{Heckler}. The final damping
of a certain scale (expressed in invariant CDM mass) from
Eq.(\ref{dampingfactor}) is 
\begin{equation}
\label{nudamping}
D = \exp{[-\left({M_{\rm D} \over M}\right)^{1/4}]},
\end{equation}
with the neutrino damping scale $M_{\rm D} \approx 1.9\times 10^{-6}
M_{\odot}$ in CDM mass. This is $3 \times 10^{-5} M_{\rm H}^{\rm CDM}$ at $T =
1.4$ MeV and corresponds to length scales $\lambda = 1/30 R_{\rm H}$.   
 
We conclude that the large peaks in the spectrum of
the radiation fluid are damped away by the time of BBN. The large peaks 
in the RAD fluid generated by the QCD transition do not give a new mechanism
for inhomogeneous BBN. 

\subsubsection{Black Hole Formation at the QCD Transition?}

It was suggested in the literature \cite{Crawford,Jedamzik,Jedamzik2} that 
the QCD transition could lead to formation of $1 M_{\odot}$ black holes, 
which could account for dark matter today. Jedamzik \cite{Jedamzik} 
proposed to identify such black holes with the MACHOs observed by 
microlensing \cite{microlensing}. He pointed out that the formation of black 
holes should be particularly efficient during the QCD epoch due to the 
significant decrease in the effective sound speed. 

In order to form a black hole in a radiation dominated 
universe, the density contrast inside the Hubble radius should be in the 
range $1/3 <\delta_{\rm H}<1$ \cite{BHF}. For an  observable amount of $1
M_{\odot}$ black holes today, i.e. $\Omega_{\rm BH}^{(0)}={\cal O} (1)$, the
fraction of energy density converted to black holes at the QCD transition must 
be ${\cal O} (a_{\rm QCD}/a_{\rm equality}) \approx 10^{-8}$. For a Gaussian
distribution this requires $\delta_{\rm rms} \approx 0.06$ (without including
any enhancement from the QCD transition) \cite{Bullock}. The QCD transition
gives an enhancement factor (at the horizon scale) of $2$ for the bag model and
of $1.5$ for lattice QCD in our linear perturbation treatment,
Figs.~\ref{fig5} - \ref{fig7}. This indicates a corresponding reduction in
the required preexisting perturbation spectrum at the solar mass scale. 
Cardall and Fuller used a qualitative argument of Carr and Hawking
\cite{CarrHawking} and the bag model and also obtained a factor $2$ reduction
in the required preexisting perturbation spectrum \cite{Cardall}. These QCD
factors of $1.5$ or $2$ are so modest that a preexisting Harrison-Zel'dovich
spectrum with COBE normalization is very far from giving a cosmologically
relevant amount of black holes \cite{COSMO}. One would have to put in a
fine-tuned tilt  $(n-1) \approx 0.4$ to get the desired amount of black
holes. However, this tilted spectrum would overproduce primordial black holes
on scales which are only a factor $50$ below the Hubble radius at the QCD
transition. Therefore a break in the preexisting spectrum below the QCD scale
would be required, a second fine tuning. 
 
We conclude that the QCD transition enhances the probability of black hole 
formation, but the preexisting spectrum needs to be fine tuned around the QCD
scale and the major effect would not be due to the QCD transition.

\subsection{Kinetic decoupling of CDM}

In Sec.~III we established the generation of peaks in the CDM spectrum
during a first-order QCD transition. Below, we discuss the properties of
the most prominent CDM candidates at temperatures of the QCD scale.
We show that weakly interacting massive particles, like the lightest 
supersymmetric particle or a heavy neutrino, are kinetically tightly coupled 
to the radiation fluid at the QCD transition and are included in the radiation
fluid in our equations. Two CDM candidates that are kinetically 
decoupled from the radiation fluid and have $p = 0$ at the QCD
epoch are the axion and primordial black holes. In our figures and equations
kinetically decoupled CDM have been labeled by CDM for brevity.
 
\subsubsection{Neutralinos and heavy neutrinos}

We start our discussion with the lightest supersymmetric particle
\cite{Griest}. In the minimal supersymmetric standard model this is the
neutralino (we assume it is stable). Constraints from LEP 2 and cosmology,
together with the assumption of universality at the GUT scale, show that its 
mass is $m_\chi > 42$ GeV \cite{Ellis}. It is essential to distinguish between
the chemical freeze-out and the kinetic decoupling of neutralinos. The
chemical freeze-out determines the amount of neutralinos today, and it happens
when the annihilation rate of neutralinos drops below the Hubble rate,
$\Gamma_{\rm ann}/H < 1$. When the neutralinos become nonrelativistic the
rate for neutralino annihilation, $\Gamma_{\rm ann} = \langle v\sigma_{\rm
ann}\rangle n_\chi$, is suppressed by the Boltzmann factor in the number
density of the neutralinos, $n_\chi \sim (m_\chi T)^{3/2}\exp(-m_\chi
/T)$. The freeze out temperature of the neutralino \cite{Griest} is $T_f \sim
m_\chi /20 > 2$ GeV, and neutralinos are chemically decoupled at the QCD transition.  

Kinetic decoupling, in contrast, is determined by the elastic scattering
between neutralinos and the dominant radiation fluid. The interaction
rate for elastic scattering is $\Gamma_{\rm el} =\tau_{\rm coll}^{-1}= \langle
v\sigma_{\rm el} \rangle n$, where $n \sim T^3$ is the number density of
relativistic particles, e.g. electrons or neutrinos. An order of magnitude
estimate shows that $\sigma_{\rm el}$ is similar to the cross section for the
elastic scattering of neutrinos and neutrons, $\sigma_{\rm el} \sim G_{\rm
F}^2 T^2$ \cite{Schramm}. We have to distinguish between the regime of perfect
kinetic coupling, i.e. neutralinos tightly coupled to the radiation fluid, an
intermediate regime where the neutralinos scatter 
elastically but the number of collisions is not sufficient to drag them along
dissipationless, and the regime of kinetic decoupling, $\Gamma_{\rm el}/H<1$,
which is roughly at $T\le 1$ MeV, since the neutralino interacts weakly.

Let us estimate the regime where the neutralinos belong to the perfect
radiation fluid. Perfectness of a fluid (dissipationless) always refers to an
external time scale, $\omega^{-1}$ of an acoustic oscillation or the Hubble
time, etc. A fluid behaves as a perfect fluid, if $\omega \tau_{\rm relax}
\ll 1$, i.e. the external time scale is larger than the relevant relaxation
time, the fluid is continually in local thermal equilibrium, and no entropy
production occurs. For the coupling of neutralinos to the radiation fluid the
relaxation time is given by $\tau_{\rm relax} = N \tau_{\rm coll}$, where $N$
is the number of collisions needed to completely change the momentum of the
neutralino due to collisions with the radiation fluid. 
The momentum transfer at a collision of a lepton with the neutralino is of
order $p_{\rm l} \sim T$. The kinetic energy of the neutralino is given by
equipartition, $p_\chi^2/2 m_\chi \sim T$, hence its momentum is 
$p_{\chi} \sim \sqrt{m_{\chi} T}$. The fractional change of the neutralino
momentum from one collision at the QCD transition is $\delta p_\chi / p_\chi
\sim \sqrt{T/m_\chi} \ll 1$. After $N$ collisions the total rms change of
momentum is $\left(\delta p_\chi/p_\chi \right)_{\rm rms} \sim \sqrt{N
T/m_\chi}$. Local thermal equilibrium is obtained if the bulk motion of the
neutralinos is governed by the leptons, i.e. the fractional change of the
neutralino momentum is of order $1$. The number of collisions needed to
completely change direction is $N \sim m_{\chi}/T \sim 300$ for $m_{\chi} =
50$ GeV and $T=T_\star$. The collision time is given by the weak interactions
rate $\Gamma_{\rm w} = \tau^{-1}_{\rm coll} \sim G_{\rm F}^2 T^5_{\star}$, and
the relaxation time is given by  
\begin{equation}
\label{turnaround}
\tau_{\rm relax}= N \tau_{\rm coll} \sim N \times 10^{-7} t_{\rm H},
\end{equation} 
with $N \sim m_{\chi}/T$. If we compare the relaxation time with the frequency of the acoustic
oscillations, we find that the condition for a perfect fluid, $\omega
\tau_{\rm relax} \ll 1$, at the QCD transition is satisfied for scales
$\lambda> \lambda_{\rm dec}(T_\star) \approx 10^{-4} R_{\rm H}$ ($m_{\chi} =
50$ GeV). Hence the neutralinos on these scales are part of the radiation
fluid at the QCD transition. Below this scale, the neutralinos cannot follow
the acoustic oscillations. On the other hand on the Hubble scale the perfect
kinetic coupling of neutralinos to the radiation fluid stops when the
required relaxation time becomes more than a Hubble time, $\tau_{\rm
relax}>t_{\rm H}$. This gives a temperature of $T_{\chi}^{\rm dec} \sim 10 $
MeV. Down to this temperature neutralinos on the Hubble scale belong to the
radiation fluid. 

Another CDM candidate is a heavy neutrino with $m_Z /2 < m_\nu < 1$ TeV 
\cite{EllisLH}, where only the upper mass bound gives cosmologically
relevant CDM. Again, the kinetic decoupling of these neutrinos happens 
way below $T_\star$. Therefore, we find that 
the neutralino or a heavy neutrino would be tightly coupled to the radiation 
fluid during the QCD transition.

\subsubsection{Axions}

One candidate for CDM at the QCD epoch is the axion \cite{rev}. We
therefore find a new mechanism to make axion clumps.

Axions could be the dominant matter today if their mass is small, i.e. 
$m_a \sim 10^{-5}$ eV, which corresponds to a breaking of the 
Peccei-Quinn (PQ) symmetry at the scale $f_{\rm PQ} \sim 10^{12}$ GeV
\cite{rev}. These axions could be produced coherently due to the initial
misalignment of the axion field and by the decay of axionic strings. The
initially  misaligned axion field starts to oscillate coherently when the
axion mass has grown to $m_a(T_1) \sim 3 H(T_1)$, where $T_1 \sim 1$ GeV
\cite{Turner86}. Thus, below $T_1 \sim 1$ GeV the oscillating axion field 
evolves as CDM, $\rho_a \propto a^{-3}$. 

In a first-order QCD transition axionic CDM falls freely into the large 
potential wells from the dominant radiation fluid. This produces peaks and
dips in the spectrum of axionic CDM. These peaks trigger the formation of 
clumps of axions with masses $10^{-20}$ -- $10^{-10} M_\odot$ shortly after
equality, see Sec.~IV.C. 

Another mechanism to produce clumps of axions has been suggested in
Ref.~\cite{Hogan}. If the reheating temperature after inflation is 
above $f_{\rm PQ}$, the axion field is inhomogeneous on scales larger than
$R_{\rm H}(T_1)$. Inhomogeneities in axion density evolve into axion 
miniclusters of typical mass $M_{\rm mc} \sim 10^{-12} M_\odot$ and 
typical radius $R_{\rm mc} \sim 10^{8}$ m $\sim 0.1 R_\odot \sim 
10^{-3}$ AU. These axion miniclusters may be enhanced by the free-fall
of matter during a first-order QCD transition. For these non-linear axion
inhomogeneities a quantitative result cannot be obtained from the linear 
analysis of Sec.~III.  Let us emphasize that, in contrast to Ref.~\cite{Hogan},
we predict axion clumps independently from the reheating temperature after 
inflation.

\subsubsection{Primordial black holes formed before the QCD transition}

A further possibility for CDM that decoupled kinetically long before the 
QCD transition is primordial black holes (PBH) produced before the QCD
transition and therefore with masses $M_{\rm BH} \ll 1 M_\odot$. In order to
survive until today PBH should have $M_{\rm BH} > 10^{15}$g 
$\approx 10^{-18} M_\odot$. PBH in the range from $10^{-18} M_\odot$ to 
$10^{-16} M_\odot$ would radiate too strongly to be compatible with 
$\gamma$-ray observations \cite{Carr}. The production of PBHs arises naturally
from hybrid inflationary scenarios \cite{inflBH,Bullock}.      

Our analysis of the QCD phase transition predicts that these small black holes 
would be attracted by the large potential wells in the radiation fluid and
would therefore tend to form miniclusters of masses up to $10^{-10} M_\odot$.
 
\subsection{Clumps in CDM}

CDM in the form of axions or PBHs is not subject to damping as 
the radiation fluid, thus the peaks in CDM will survive until structure
formation starts. The free streaming scale of CDM is way below our smallest
scales, because the initial velocity of axions or PBHs is completely
negligible. An overdensity of CDM in the form of axions or PBH (or any
other matter that is kinetically decoupled at the QCD epoch) decouples from
the cosmic expansion when its density contrast becomes nonlinear,
$(\delta\rho/\rho)_R \sim 1$ (condition for turn around). It collapses and
virializes by violent gravitational relaxations and forms a clump of CDM. For
spherical collapse the final viral radius is half of the radius at turn around
\cite{Padmanabhan}.  

We take a COBE \cite{COBE} normalized spectrum and allow for a tilt $|n-1| \le
0.2$. During the radiation dominated regime, $(\delta \rho/\rho)_k$ for CDM
continues to grow logarithmically. In $(\delta\rho/\rho)^2_R$ another
logarithm comes from summing modes up to $k \sim 1/R$, where $R$ is the radius
of the window function. The enhancement factor ${\cal E}$ of CDM density
fluctuations during the QCD transition has been obtained in
Eq.~(\ref{enhancement}). Putting these factors together we obtain the
amplitude of CDM perturbations of size $R$ at the time of equality: 
\begin{equation}
\left(\delta\rho \over \rho\right)^{\rm CDM}_R (t_{\rm eq}) \approx 
2 \times 10^{-4} \left( {k\over k_0} \right)^{n-1\over 2} 
\left[\ln({k\over k_{\rm eq}})\right]^{\frac32}{\cal E}_k(t_{\rm eq})
\end{equation}
$k_0$ denotes the comoving wave number of the mode crossing the horizon 
today. In the following we assume $\Omega_0 h^2 = 1/4$, and thus 
$z_{\rm eq} \approx 6000$. For a CDM mass of $10^{-15}(10^{-20}) M_\odot$ a 
tilt of $n - 1 = 0.2$ gives a factor $\approx 16(23)$. The logarithms
contribute another factor $\approx 94(140)$. The enhancement factor has been
calculated in Sec.~III to be ${\cal E}(t_{\rm eq}) \approx 2(8)$ for the
lattice QCD fit. Looking at $3\sigma$ peaks we find without tilt ($n-1=0$)
\begin{equation}
\left(\delta\rho \over \rho\right)^{3 \sigma, n=1}_R (t_{\rm eq}) \approx
0.1(0.6) \quad \mbox{for\ } M_{\rm clump} = 10^{-15} (10^{-20}) M_\odot \ .
\end{equation}
This implies that these clumps go nonlinear at $z_{\rm nl} 
\approx 600 (3600)$ and collapse to clumps of radius 
$R_{\rm clump} \approx 14 (0.04) R_\odot$. With tilt ($n-1=0.2$) we find
\begin{equation}
\left(\delta\rho \over \rho\right)^{3 \sigma, n=1.2}_R (t_{\rm eq}) \approx
2 (16) \quad \mbox{for\ } M_{\rm clump} = 10^{-15} (10^{-20}) M_\odot \ .
\end{equation}
These over-densities start to collapse even before $t_{\rm eq}$ because locally
CDM starts to dominate at $\sim 2(16) T_{\rm eq}$. This leads to clump 
radii of $R_{\rm clump} \approx 0.7 (0.003) R_\odot$. 

We conclude that the peaks in the CDM spectrum lead to clumps 
of masses $10^{-20} - 10^{-10} M_\odot$. Today, these clumps would 
have a density contrast of $10^{10} - 10^{17}$, where the lower value
corresponds to a $10^{-15} M_\odot$ clump from an untilted CDM spectrum, the
bigger value is for a $10^{-20} M_\odot$ clump from a tilted CDM spectrum. 
The evolution of these clumps in the late stages of structure formation
remains to be investigated (disruption, mergers, etc.). 

For larger enhancement, e.g., if it should turn out that the latent heat
is bigger than the value from present lattice QCD calculations, more compact
clumps are possible. These could be subject to femto-lensing \cite{FL}. 
With the values of our lattice fit, the CDM clumps are not
compact enough to lie within the Einstein radius, which is  $R_{\rm E} 
\sim 0.02 R_\odot$ for a $10^{-15} M_\odot$ clump.

The clumping of CDM changes the
expected reaction rates for some dark matter searches,  
because some of the rates depend on the space-time position of the detector, 
star, or planet. Especially experiments looking for axion decay in 
strong magnetic fields \cite{Sikivie,rev} would not yield limits on the
axion mass. Maybe these experiments just tell us that we are not sitting in an 
axion cloud currently. These consequences remain to be studied further. 

\acknowledgements
We are very grateful to U. Heller and F. Karsch for providing us with 
their lattice QCD data sets. We like to thank J. A. Bardeen, V. Berezinsky,
B. J. Carr, U. Heller, J. Ignatius, P. Jetzer, F. Karsch, H. Markum, 
V. Mukhanov, J. Silk, P. Sikivie, and N. Straumann for helpful discussions 
and references to the literature. C.S. thanks J. Silk and the
Center for Particle Astrophysics in Berkeley for hospitality.
D.J.S. and P.W. thank the Swiss National Science Foundation for financial
support. D.J.S. is an Alexander von Humboldt fellow.

\end{document}